\documentclass[author]{aa}  
\begin{document}

   \title{Linking interstellar and cometary O$_2$:
a deep search for $^{16}$O$^{18}$O in the solar-type protostar IRAS 16293--2422} 

\titlerunning{Linking interstellar and cometary O$_2$}

\author{V. Taquet\inst{1} 
\and E.~F. van Dishoeck\inst{2,3} 
\and M. Swayne\inst{2} 
\and D. Harsono\inst{2} 
\and J.~K. J{\o}rgensen\inst{4} 
\and L. Maud\inst{2} 
\and N.~F.~W. Ligterink\inst{2} 
\and H.~S.~P. M{\"u}ller\inst{5} 
\and C. Codella\inst{1} 
\and K. Altwegg\inst{6} 
\and A. Bieler\inst{7} 
\and A. Coutens\inst{8} 
\and M.~N. Drozdovskaya\inst{9} 
\and K. Furuya\inst{10} 
\and M.~V.~Persson\inst{11} 
\and M.~L.~R. van 't Hoff\inst{2} 
\and C. Walsh\inst{12} 
\and S.~F. Wampfler\inst{9} }

\institute{INAF, Osservatorio Astrofisico di Arcetri, Largo E. Fermi 5, 50125 Firenze, Italy 
\email{taquet@arcetri.astro.it}
\and Leiden Observatory, Leiden University, P.~O.~Box 9531, 2300~RA Leiden, The Netherlands 
\and Max-Planck-Institut f\"{u}r extraterretrische Physik, Giessenbachstrasse 1, 85748 Garching, Germany
\and Centre for Star and Planet Formation, Niels Bohr Institute \& Natural History Museum of Denmark, University of Copenhagen, {\O}ster Voldgade 5–7, DK-1350 Copenhagen K., Denmark
\and I. Physikalisches Institut, Universit{\"a}t zu K{\"o}ln, Z{\"u}lpicher Str. 77, 50937 K{\"o}ln, Germany
\and Physikalisches Institut, Universit{\"a}t Bern, Sidlerstrasse 5, 3012 Bern, Switzerland
\and Climate and Space Sciences and Engineering, University of Michigan, Ann Arbor, MI 48109, USA
\and Laboratoire d'Astrophysique de Bordeaux, Univ. Bordeaux, CNRS, B18N, allée Geoffroy Saint-Hilaire, 33615 Pessac, France
\and Center for Space and Habitability, University of Bern, Gesellschaftsstrasse 6, CH-3012 Bern, Switzerland
\and Center for Computer Sciences, University of Tsukuba, 305-8577 Tsukuba, Japan
\and Department of Space, Earth, and Environment, Chalmers University of Technology, Onsala Space Observatory, 439 92 Onsala, Sweden
\and School of Physics and Astronomy, University of Leeds, Leeds LS2 9JT, UK
}

   \date{}

 
  \abstract
   {
Recent measurements carried out at comet 67P/Churyumov–Gerasimenko (67P/C-G) with the {\it Rosetta} probe revealed that molecular oxygen, O$_2$, is the fourth most abundant molecule in comets. 
Models show that O$_2$ is likely of primordial nature, coming from the interstellar cloud from which our Solar System was formed. 
However, gaseous O$_2$ is an elusive molecule in the interstellar medium with only one detection towards quiescent molecular clouds, in the $\rho$ Oph A core. 
We perform a deep search for molecular oxygen, through the $2_1 - 0_1$ rotational transition at 234 GHz of its $^{16}$O$^{18}$O isotopologue, towards the warm compact gas surrounding the nearby Class 0 protostar IRAS 16293--2422 B with the ALMA interferometer. 
We also look for the chemical daughters of O$_2$, HO$_2$ and H$_2$O$_2$. Unfortunately, the H$_2$O$_2$ rotational transition is dominated by ethylene oxide c-C$_2$H$_4$O while HO$_2$ is not detected. 
The targeted $^{16}$O$^{18}$O transition is surrounded by two brighter transitions at $\pm 1$ km s$^{-1}$ relative to the expected $^{16}$O$^{18}$O transition frequency. 
After subtraction of these two transitions, residual emission at a 3$\sigma$ level remains, but with a velocity offset of $0.3 - 0.5$ km s$^{-1}$ relative to the source velocity, rendering the detection "tentative".
We derive the O$_2$ column density for two excitation temperatures $T_{\rm ex}$ of 125 and 300 K, as indicated by other molecules, in order to compare the O$_2$ abundance between IRAS16293 and comet 67P/C-G. 
Assuming that $^{16}$O$^{18}$O is not detected and using methanol CH$_3$OH as a reference species, we obtain a [O$_2$]/[CH$_3$OH] abundance ratio lower than $2-5$, depending on the assumed $T_{\rm ex}$, a three to four times lower abundance than the [O$_2$]/[CH$_3$OH] ratio of $5-15$ found in comet 67P/C-G.
Such a low O$_2$ abundance could be explained by the lower temperature of the dense cloud precursor of IRAS16293 with respect to the one at the origin of our Solar System that prevented an efficient formation of O$_2$ in interstellar ices.
}


   \maketitle
%

\section{Introduction}

Molecular oxygen O$_2$ has recently been detected in surprisingly large quantities towards Solar System comets. \citet{Bieler2015} first detected O$_2$ in comet 67P/Churyumov– Gerasimenko (hereinafter 67P/C-G) with the mass spectrometer ROSINA ("Rosetta Orbiter Spectrometer for Ion and Neutral Analysis") on the {\it Rosetta} probe and derived an averaged high abundance of $3.80 \pm 0.85$ \% relative to water. This surprising detection has since been confirmed by UV spectroscopy in absorption by \citet{Keeney2017} using the Alice far-ultraviolet spectrograph with an even higher abundance of $11-68$ \% (with a median value of 25 \%). 
A re-analysis of the data from the Neutral Mass Spectrometer on board the {\it Giotto} mission which did a fly-by of comet 1P/Halley in 1986 allowed \citet{Rubin2015b} to confirm the presence of O$_2$ at similar levels to that seen in comet 67P/C-G by ROSINA. 
All these detections therefore suggest that O$_2$ should be abundantly present in both Jupiter-family comets, such as 67P/C-G, and Oort Cloud comets, such as 1P/Halley, which have different dynamical behaviours and histories.

The ROSINA instrument not only revealed a high abundant of molecular oxygen but also that the O$_2$ signal is strongly correlated with water, unlike other di-atomic species with similar volatilities such as N$_2$ or CO \citep{Bieler2015, Rubin2015a}. 
\citet{Bieler2015} therefore claimed that gas phase chemistry is not responsible for the detection of O$_2$. Instead, the detected O$_2$ should come from the sublimation of O$_2$ ice trapped within the bulk H$_2$O ice matrix suggesting that O$_2$ was present in ice mantles before the formation of comet 67P/C-G in the presolar nebula.  
Several explanations have been suggested to explain the presence of O$_2$ in comets. 
\citet{Taquet2016} explored different scenarios to explain the high abundance of O$_2$, its strong correlation with water, and the low abundance of the chemically related species H$_2$O$_2$, HO$_2$, and O$_3$. They show that a formation of solid O$_2$ together with water through surface chemistry in a dense (i.e. $n_{\rm H} \sim 10^6$ cm$^{-3}$) and relatively warm ($T \sim 20$ K) dark cloud followed by the survival of this O$_2$-H$_2$O ice matrix in the pre-solar and solar nebulae could explain all the constraints given by {\it Rosetta}. Such elevated temperatures are needed to enhance the surface diffusion of O atoms that recombine to form solid O$_2$ and to limit the lifetime of atomic H on grains and prevent the hydrogenation of O$_2$. 
\citet{Mousis2016} developed a toy model in which O$_2$ is only formed through the radiolysis of H$_2$O, and showed that O$_2$ can be formed in high abundances (i.e. [O$_2$]/[H$_2$O] $\geq 1$\%) in dark clouds. However, laboratory experiments demonstrate that the production of O$_2$ through radiolysis should be accompanied by an even more efficient production of H$_2$O$_2$ \citep{Zheng2006} contradicting the low [H$_2$O$_2$]/[O$_2$] abundance ratio of $(0.6 \pm 0.07) \times 10^{-3}$ measured by {\it Rosetta} in comet 67P/C-G. 
\citet{Dulieu2017} experimentally showed that O$_2$ can be produced during the evaporation of a H$_2$O-H$_2$O$_2$ ice mixture through the dismutation of H$_2$O$_2$. However, although O$_2$ is produced in large quantities in these experiments, the dismutation is not efficient enough to explain the low abundance of H$_2$O$_2$ relative to O$_2$ measured by \citet{Bieler2015}. 

If the \citet{Taquet2016} explanation holds, O$_2$ should be detectable in molecular clouds. 
However, O$_2$ is known to be an elusive molecule in the interstellar medium. Recent high sensitivity observations with the {\it Herschel Space Observatory} allowed for deep searches of O$_2$ in dark clouds and Solar System progenitors. 
O$_2$ has been detected towards only two sources: the massive Orion star-forming region \citep[O$_2$/H$_2$ $\sim 0.3-7.3 \times 10^{-6}$;][]{Goldsmith2011, Chen2014} and the low-mass dense core $\rho$ Oph A located in the Ophiucus molecular cloud \citep[O$_2$/H$_2$ $\sim 5 \times 10^{-8}$;][]{Larsson2007, Liseau2012}. Interestingly, with a high density $n_{\rm H}$ of $\sim 10^6$ cm$^{-3}$ and a warm temperature $T$ of $\sim 24-30$ K, $\rho$ Oph A presents exactly the physical conditions invoked by \citet{Taquet2016} to trigger an efficient formation of O$_2$ in ices. 

However, O$_2$ has yet to be found in Solar System progenitors. A deep search for O$_2$ towards the low-mass protostar NGC1333-IRAS4A located in the Perseus molecular cloud by \citet{Yildiz2013} using {\it Herschel} resulted in an upper limit only on the O$_2$ abundance ([O$_2$]/[H$_2$] $< 6 \times 10^{-9}$).
The search for O$_2$ towards NGC1333-IRAS4A using {\it Herschel} suffered from a high beam dilution due to the large beam of the telescope at the frequency of the targeted O$_2$ transition (44$\arcsec$ at 487 GHz) with respect to the expected emission size (a few arcsec). 
In addition, NGC1333-IRAS4A is located in the relatively cold Perseus molecular cloud. Dust temperature maps of Perseus obtained from PACS and SPIRE observations using {\it Herschel} as part of the Gould Belt survey \citep{Andre2010} suggest a dust temperature of $\sim 13-14$ K in the NGC1333 star-forming region.

In this work, we present deep high angular resolution observations of $^{16}$O$^{18}$O towards the brightest low-mass binary protostellar system IRAS 16293-2422 (hereinafter IRAS16293) with the Atacama Large Millimeter/submillimeter Array (ALMA). 
As the main isotopologue of molecular oxygen is almost unobservable from the ground due to atmospheric absorption, we targeted its $^{16}$O$^{18}$O isotopologue through its $2_1 - 0_1$ rotational transition at 233.946 GHz ($E_{\rm up} = 11.2$ K, $A_{\rm i,j} = 1.3 \times 10^{-8}$ s$^{-1}$). 
The angular resolution is about 0\farcs5, which is comparable to the emission size of most molecular transitions observed towards the binary system \citep{Baryshev2015, Jorgensen2016}.
We also targeted transitions from the chemical "daughter" species of O$_2$, HO$_2$, and H$_2$O$_2$, thought to be formed at the surface of interstellar ices through hydrogenation of O$_2$. 
In addition to being closer \citep[141 vs 235 pc;][]{Hirota2008, OrtizLeon2017} and more luminous \citep[21 vs 9.1 $L_{\odot}$;][]{Jorgensen2005, Karska2013} than NGC1333-IRAS4A, IRAS16293 is located in the same molecular cloud as $\rho$ Oph A, Ophiuchus. IRAS16293 is therefore located in a slightly warmer environment with a dust temperature of $\sim 16$ K in its surrounding cloud (B. Ladjelate, private communication), favouring the production of O$_2$ in ices according to the scenario presented by \citet{Taquet2016}.

\section{Observations and data reduction}

IRAS16293, located at 141 pc, has a total luminosity of 21 $L_{\odot}$ and a total envelope mass of 2 $M_{\odot}$ \citep{Jorgensen2005, Lombardi2008, OrtizLeon2017, Dzib2018}. It consists of a binary system with two sources A and B separated by 5.1$\arcsec$ or 720 AU \citep{Looney2000, Chandler2005}. 
Due to its bright molecular emission and relatively narrow transitions, IRAS16293 has been a template for astrochemical studies \citep[see][for a more detailed overview of the system]{Jorgensen2016}. 
Source A, located towards the South-East of the system, has broader lines than source B that could possibly be attributed to the different geometries of their disks. Transitions towards Source A present a velocity gradient consistent with the Keplerian rotation of an inclined disk-like structure whereas Source B is close to be face-on \citep{Pineda2012, Zapata2013}.
Several unbiased chemical surveys have been carried out towards IRAS16293 using single-dish or interferometric facilities \citep{Caux2011, Jorgensen2011} to obtain a chemical census of this source. 
A deep ALMA unbiased chemical survey of the entire Band 7 atmospheric window between 329.15 and 362.90 GHz has recently been performed in the framework of the Protostellar Interferometric Line Survey \citep[PILS;][]{Jorgensen2016}.
The unprecendented sensitivity and angular resolution offered by ALMA allows to put strong constraints on the chemical organic composition and the physical structure of the protostellar system \citep{Jorgensen2016, Jorgensen2018, Coutens2016, Coutens2018, Lykke2017, Ligterink2017, Jacobsen2018, Persson2018, Drozdovskaya2018}. 
{  The $^{16}$O$^{18}$O $3_2-1_1$ transition at 345.017 GHz lies in the ALMA PILS frequency range. However, this line is expected to be much weaker than the $2_1-0_1$ transition at 233.946 GHz due to its lower Einstein coefficient ($1.8 \times 10^{-9}$ vs $1.3 \times 10^{-8}$ s$^{-1}$). A simple model assuming Local Thermal Equilibrium and an excitation temperature $T_{\rm ex}$ of 300 K suggests that the intensity of the 345.017 GHz transition is five times lower than that at 233.946 GHz, suggesting that it cannot provide deeper constraints on the O$_2$ column density towards IRAS16293.}

IRAS16293 was observed with the 12m antenna array of ALMA during Cycle 4, under program 2016.1.01150.S (PI: Taquet), with the goal of searching for $^{16}$O$^{18}$O at a similar angular resolution as the PILS data. The observations were carried out during four execution on 2016 November 10, 20, 22, and 26 in dual-polarization mode in Band 6.
IRAS 16293 was observed with one pointing centered on $\alpha_{\rm J2000}$ = 16:32:22.72, $\delta_{\rm J2000}$ = -24:28:34.3 located between sources A and B.
39-40 antennas of the main array were used, with baselines ranging from 15.1 to 1062.5 m. The primary beam is 25\farcs6 while the synthesized beam has been defined to 0\farcs5 to match the beam size of the PILS data. 
The bandpass calibrators were J1527-2422 (execution 1) and J1517-2422 (executions 2 to 4), the phase calibrator was J1625-2527, and the flux calibrators were J1527-2422 (executions 1 to 3) and J1517-2422 (execution 4).
Four spectral windows were observed each with a bandwidth of 468.500 MHz and a spectral resolution of 122 kHz or 0.156 km s$^{-1}$ and covered $233.712 - 234.180$, $234.918-235.385$, $235.908-236.379$, and $236.368-236.841$ GHz. 
The data were calibrated with the CASA software \citep[version 4.7.3]{McMullin2007}. 

The continuum emission has been subtracted from the original datacube in order to image individual transitions. Due to the high sensitivity of the data, it is impossible to find spectral regions with line-free channels that can be used to derive the continuum emission. Instead, we follow the methodology defined in \citet{Jorgensen2016} to obtain the continuum emission maps that can be used to subtract it from the original datacubes.  
In short, the continuum is determined in two steps. First, a Gaussian function is used to fit the emission distribution towards each pixel of the datacube. A second Gaussian function is then fitted to the part of the distribution within $F \pm \Delta F$ where $F$ and $\Delta F$ are the centroid and the width of the first Gaussian, respectively. The centroid of the second Gaussian function is then considered as the continuum level for each pixel. 

After the continuum subtraction, the four final spectral line datacubes have a rms sensitivity of 1.2 - 1.4 mJy beam$^{-1}$ channel$^{-1}$ or 0.47 - 0.55 mJy beam$^{-1}$ km s$^{-1}$. This provides the deepest ALMA dataset towards this source in this Band obtained so far.

\begin{table}[htp]
\centering
\caption{Properties of the transitions targeted in this work.} 
\begin{tabular}{l c c c c}
\hline							
\hline							
Species	&	Transition	&	Frequency	&	$A_{\rm i,j}$	&	$E_{\rm up}$	\\
	&		&	(GHz)	&	(s$^{-1}$)	&	(K)	\\
  \hline
$^{16}$O$^{18}$O & $2_1$-$0_1$ & 233.94610  & 1.3(-8)  & 11.2 \\
\hline                  
HO$_2$  & $4_{1,4,9/2,5}$-$5_{0,5,9/2,5}$  & 235.14215  & 4.3(-6)  & 58.2 \\
HO$_2$  & $4_{1,4,9/2,4}$-$5_{0,5,9/2,4}$  & 235.16902  & 4.3(-6) & 58.2 \\
HO$_2$  & $4_{1,4,9/2,5}$-$5_{0,5,11/2,5}$ & 236.26779  & 1.3(-6) & 58.2 \\
HO$_2$  & $4_{1,4,9/2,5}$-$5_{0,5,11/2,6}$ & 236.28092  & 7.7(-5) & 58.2 \\
HO$_2$  & $4_{1,4,9/2,4}$-$5_{0,5,11/2,5}$ & 236.28442  & 7.8(-5) & 58.2 \\
\hline                  
H$_2$O$_2$  & $4_{2,3}$-$5_{1,5}$ & 235.95594  & 5.0(-5) & 77.6 \\
\hline									
\end{tabular}
\label{transitions}
\tablebib{
The $^{16}$O$^{18}$O, HO$_2$, H$_2$O$_2$ spectroscopic data used in this work are from \citet{Drouin2010}, \citet{Chance1995}, and \citet{Petkie1995}, respectively.
}
\end{table}

\section{Results}

\subsection{Overview of the Band 6 data} \label{overview_spectrum}

As discussed by \citet{Lykke2017}, around source B, most molecular transitions reach their intensity maximum about 0\farcs25 away from the continuum peak of IRAS16293 B in the south-west direction \citep[see also the images in][]{Baryshev2015}. However, most transitions towards this position are usually optically thick and absorption features are prominent. A "full-beam" offset position, located twice further away relative to the continuum peak, in the same direction gives a better balance between molecular emission intensities and absorption features.
Figures \ref{spect_spw1} to \ref{spect_spw4} of the Appendix show the spectra of the four spectral windows obtained towards the full-beam offset position located 0.5$\arcsec$ away from the continuum peak of IRAS16293 B in the south-west direction whose coordinates are $\alpha_{\rm J2000}$ = 16:32:22.58, $\delta_{\rm J2000}$ = -24:28:32.8. 

We detect a total of 671 transitions above the 5$\sigma$ level, giving a line density of 358 transitions per GHz or one transition every 2.8 MHz. 
To identify the transitions, a Local Thermal Equilibrium (LTE) model is applied assuming a systemic velocity $V_{\rm lsr} = 2.7$ km s$^{-1}$, a linewidth $FWHM = 1$ km s$^{-1}$, {  and a Gaussian source distribution with a size of 0\farcs5, resulting in a beam filling factor of 0.5}. We used the column densities and excitation temperatures derived from the PILS survey \citep{Jorgensen2016, Jorgensen2018, Coutens2016, Lykke2017, Ligterink2017, Fayolle2017, Persson2018, Drozdovskaya2018}.
The model computes the intensity following the methodology summarised in \citet{Goldsmith1999}. In particular, the overall opacity of each transition is computed but does not affect the profile of the transition that is assumed to remain Gaussian.
A total number of 253 spectroscopic entries mostly using the CDMS \citep{Muller2005, Endres2016} and JPL \citep{Pickett1998} catalogues including rare isotopologues and vibrationally excited states have been used. 
The LTE model overestimates the intensity of most common species since the optical depths still remain high even at a distance of 0\farcs5 away from the continuum peak.
In spite of the high number of species included in the model, $\sim 70$ \% of the $\sim 670$ transitions remain unidentified at a 5$\sigma$ level. 
Most identified transitions are attributed to oxygen-bearing complex organic molecules including their main and rare isotopologues, such as methyl formate CH$_3$OCHO, acetic acid CH$_3$COOH, acetaldehyde CH$_3$CHO, ethylene glycol (CH$_2$OH)$_2$, ethanol C$_2$H$_5$OH, or methanol CH$_3$OH. 
This serves as warning that care must be taken with identifications based on single line.
%
A significant part of unidentified lines could be due to additional vibrationally excited states and isotopologues of COMs that are not yet characterised by spectroscopists.



\subsection{Analysis of the $^{16}$O$^{18}$O transition}

\begin{figure*}[htp]
\centering 
\includegraphics[height=0.51\columnwidth]{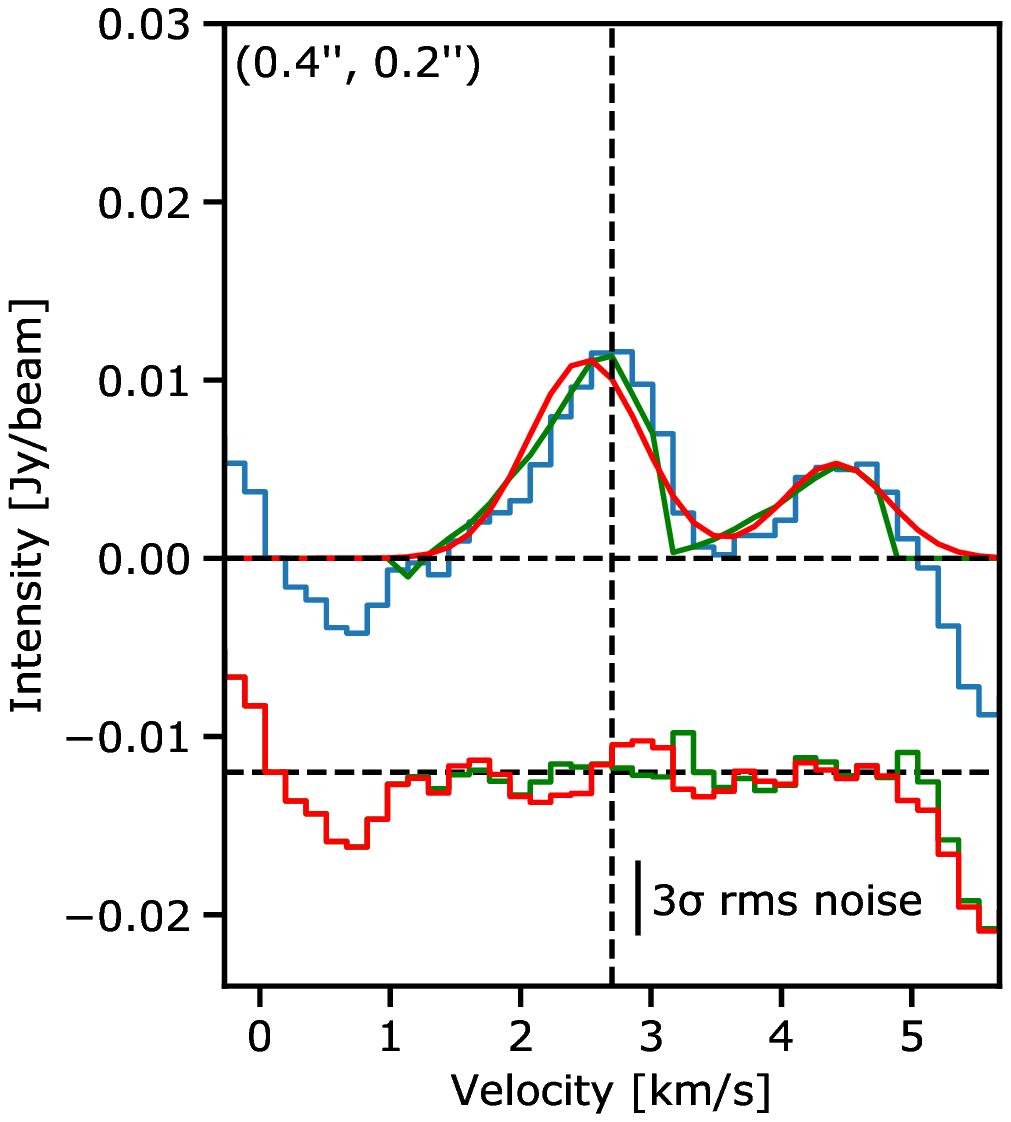} 
\includegraphics[height=0.51\columnwidth]{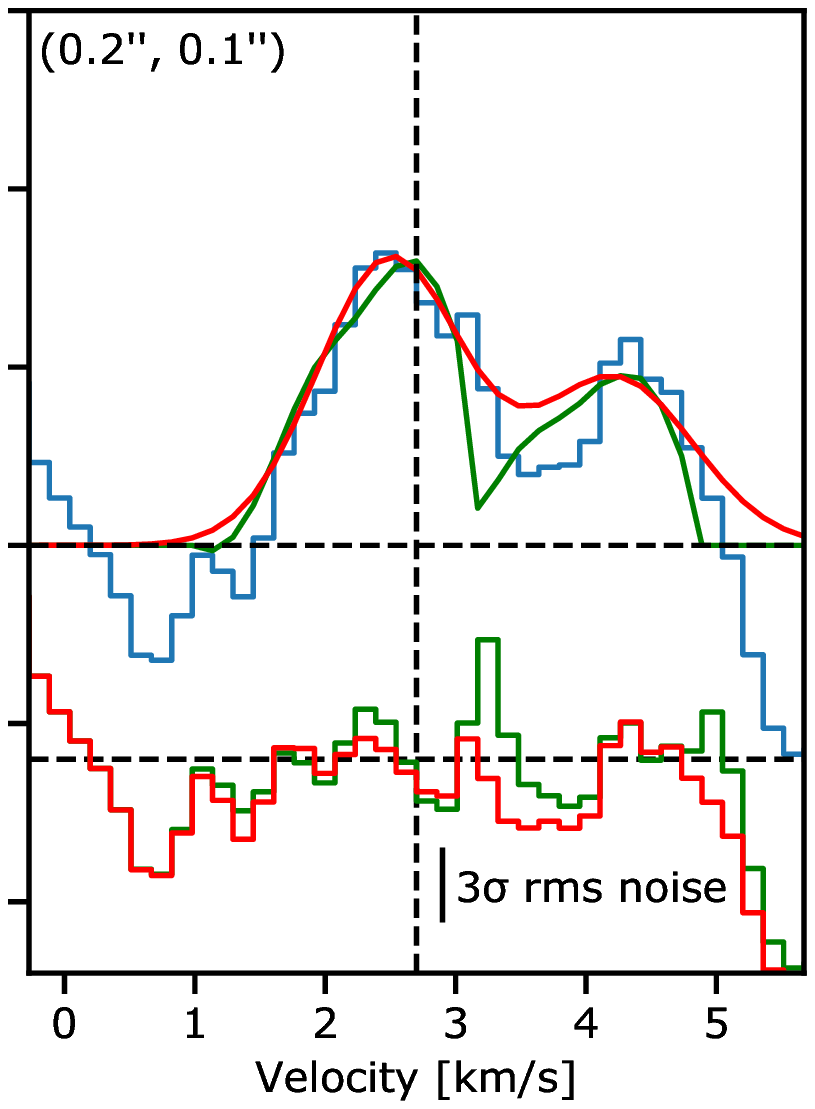} 
\includegraphics[height=0.51\columnwidth]{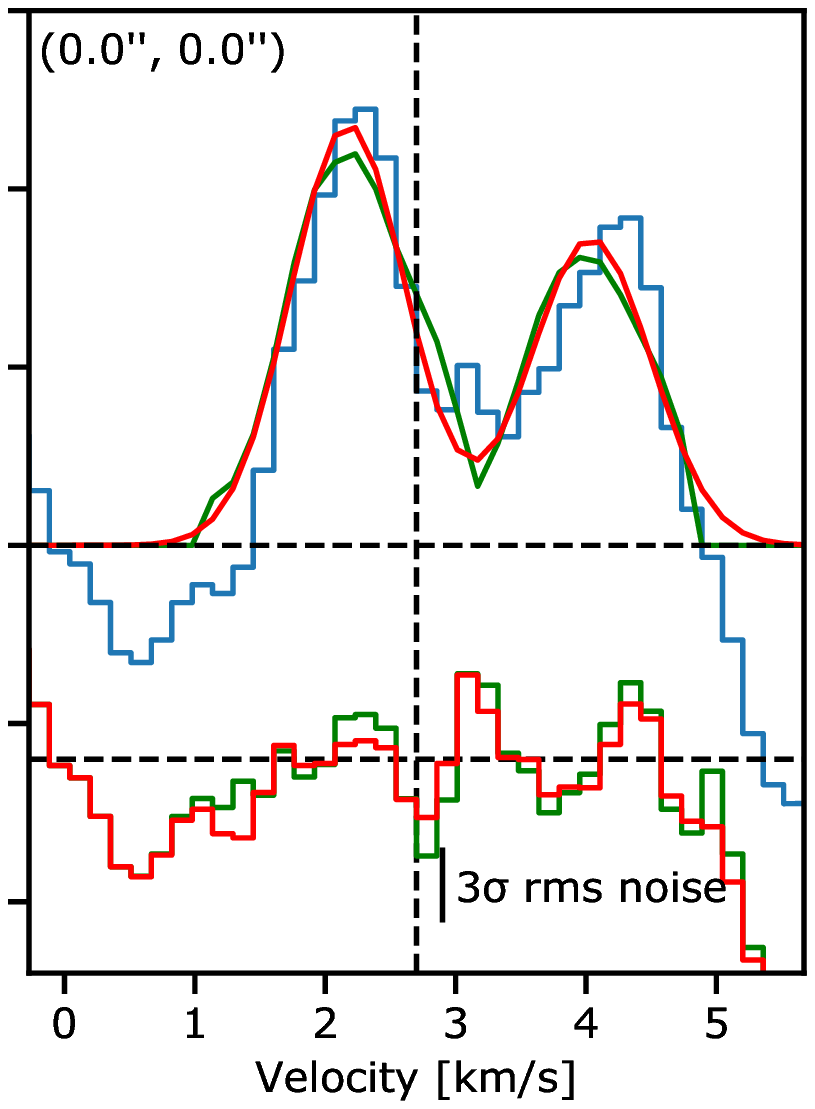}
\includegraphics[height=0.51\columnwidth]{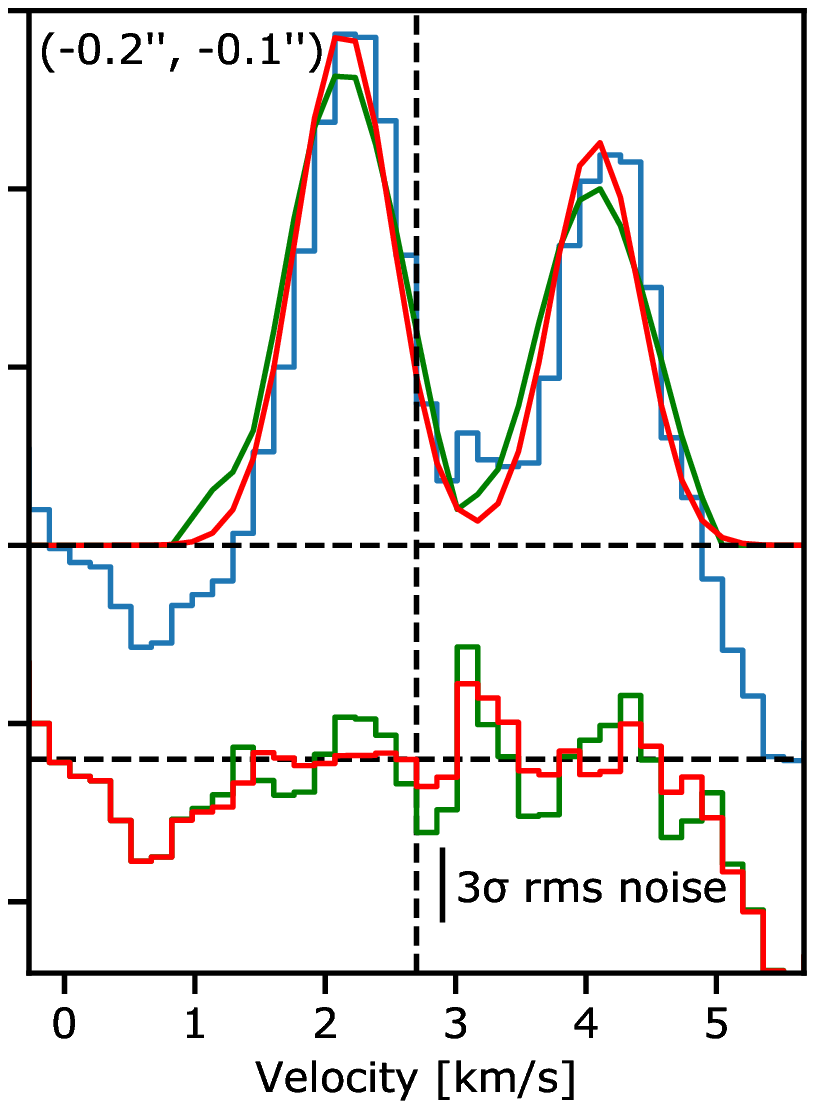} 
\includegraphics[height=0.51\columnwidth]{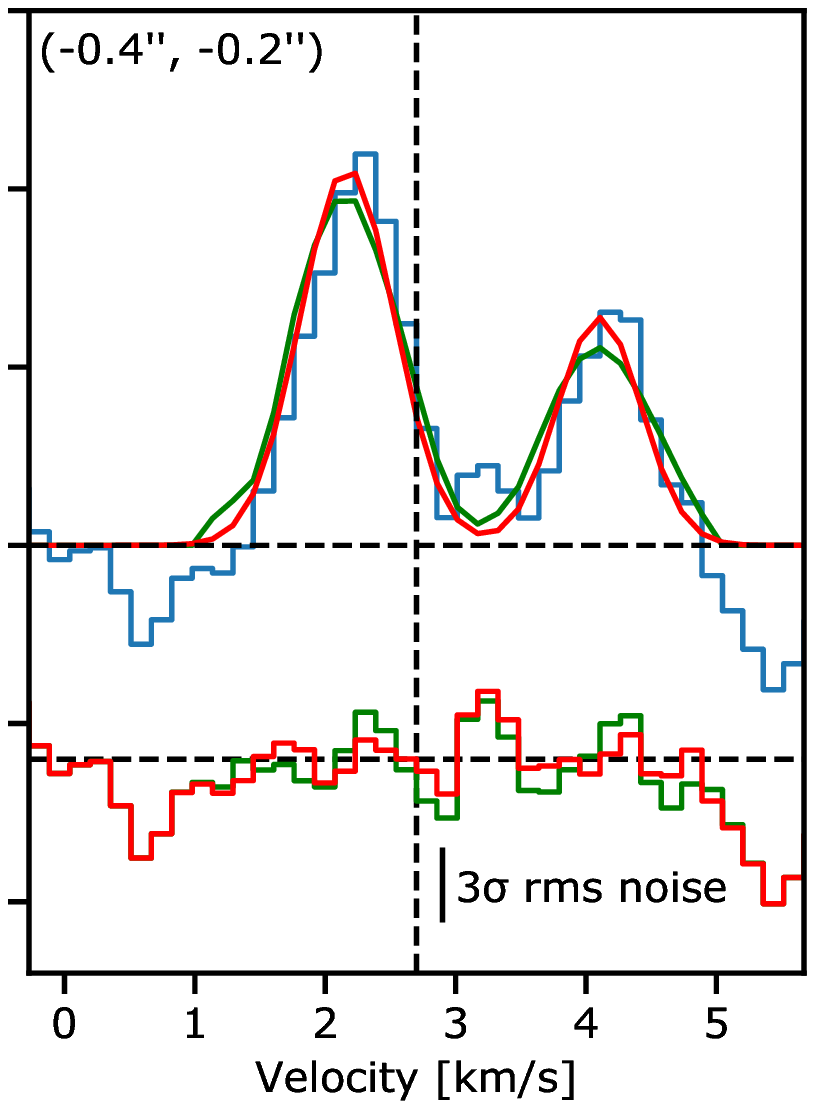} 
\caption{Observed spectra (in blue) around the $^{16}$O$^{18}$O transition at 233.946098 GHz towards five positions in IRAS16293 B depicted in the maps in Figure \ref{maps_o2}. 
Red and green lines show the best-fit curves to the data assuming a Gaussian profile and the profile from the CH$_3$NCO transition at 234.08809 GHz, respectively. 
Residual spectra of the best-fit Gaussian and reference transition profiles are shown below the observed spectra in red and green, respectively.
Positions in arcsec relative to the continuum peak position of IRAS16293 B are shown at the top left of each panel. 
The vertical and horizontal black dashed lines depict the source velocity at 2.7 km s$^{-1}$ and the baseline, respectively.
}
\label{spect_o2}
\end{figure*}

Figure \ref{spect_o2} shows the spectra around the $^{16}$O$^{18}$O transition at 233.946 GHz towards the continuum peak of IRAS16293 B, with a source velocity $V_{\rm LSR} = 2.7$ km s$^{-1}$, as well as the half-beam and full-beam offset positions in the North-East and South-West directions. 
It can be seen that the $^{16}$O$^{18}$O transition is surrounded by two brighter transitions peaking at $1.9$ km s$^{-1}$ and 3.8 km s$^{-1}$. 
Line identification analysis using the CDMS and JPL databases only revealed one possibility, the hydroxymethyl CH$_2$OH radical whose laboratory millimeter spectrum has recently been obtained by \citet{Bermudez2017}. In spite of the high uncertainty of 4 MHz for the frequency of these two transitions, this doublet is the best match with a doublet splitting frequency of 1.56 MHz (= 2.0 km s$^{-1}$), and with similar upper level energies of 190 K and Einstein coefficients of $3.7 \times 10^{-5}$ s$^{-1}$. 
We fitted the doublet around 233.946 GHz by varying the CH$_2$OH column density and assuming an excitation temperature $T_{\rm ex} = 125$ K, following the excitation temperature found for complex organic molecules by \citet{Jorgensen2016}. The doublet towards the western half-beam position, whose coordinates are (-0.2$\arcsec$; -0.1$\arcsec$) relative to IRAS16293 B, is best fitted with $N$(CH$_2$OH) = $3 \times 10^{16}$ cm$^{-2}$ = 0.15 \% relative to CH$_3$OH, assuming $N$(CH$_3$OH) = $2 \times 10^{19}$ cm$^{-2}$ \citep{Jorgensen2016}. Using the derived column density and an excitation temperature of 125 K, CH$_2$OH should have several detectable, free from contamination, transitions in the PILS data at Band 7. However, the four brightest transitions at $\sim 362$ GHz with $E_{\rm up}$ of 223 K are expected to have intensity peaks of $\sim 0.1$ Jy beam$^{-1}$ but are not detected with a sensitivity of 5 mJy beam km s$^{-1}$, a factor of 20 lower, negating this identification of CH$_2$OH. 

As shown by \citet{Pagani2017} who searched for $^{16}$O$^{18}$O towards Orion KL with ALMA, the transition at $\sim 4$ km s$^{-1}$ shown in Fig. \ref{spect_o2} could be attributed to two transitions at 233.944899 and 233.945119 GHz (or 3.9 and 4.2 km s$^{-1}$ ) from the vibrationally excited state $v_{13}+v_{21}=1$ of C$_2$H$_5$CN.

\begin{figure*}[htp]
\centering 
\includegraphics[width=0.6\columnwidth]{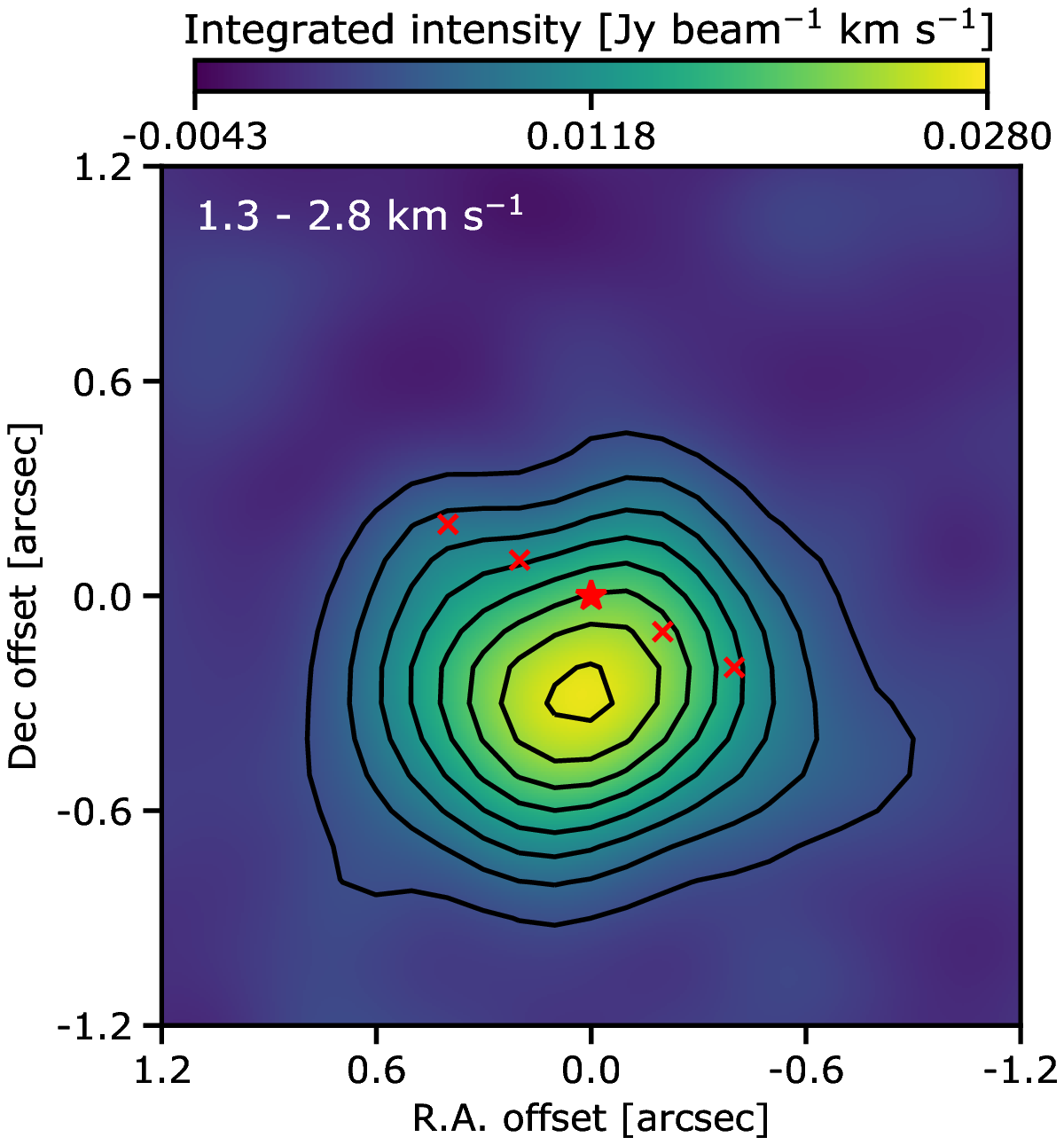} 
\includegraphics[width=0.6\columnwidth]{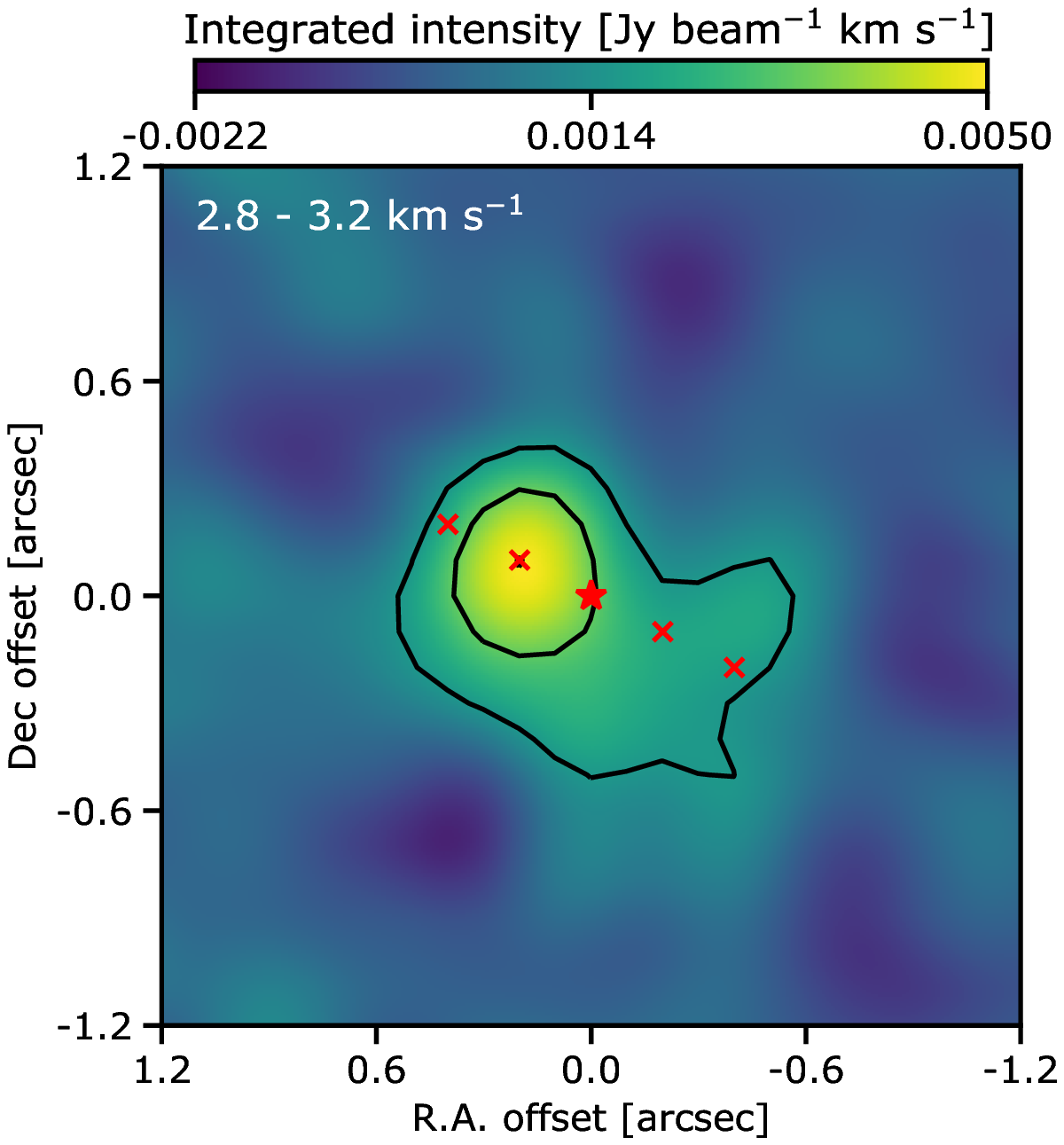} 
\includegraphics[width=0.6\columnwidth]{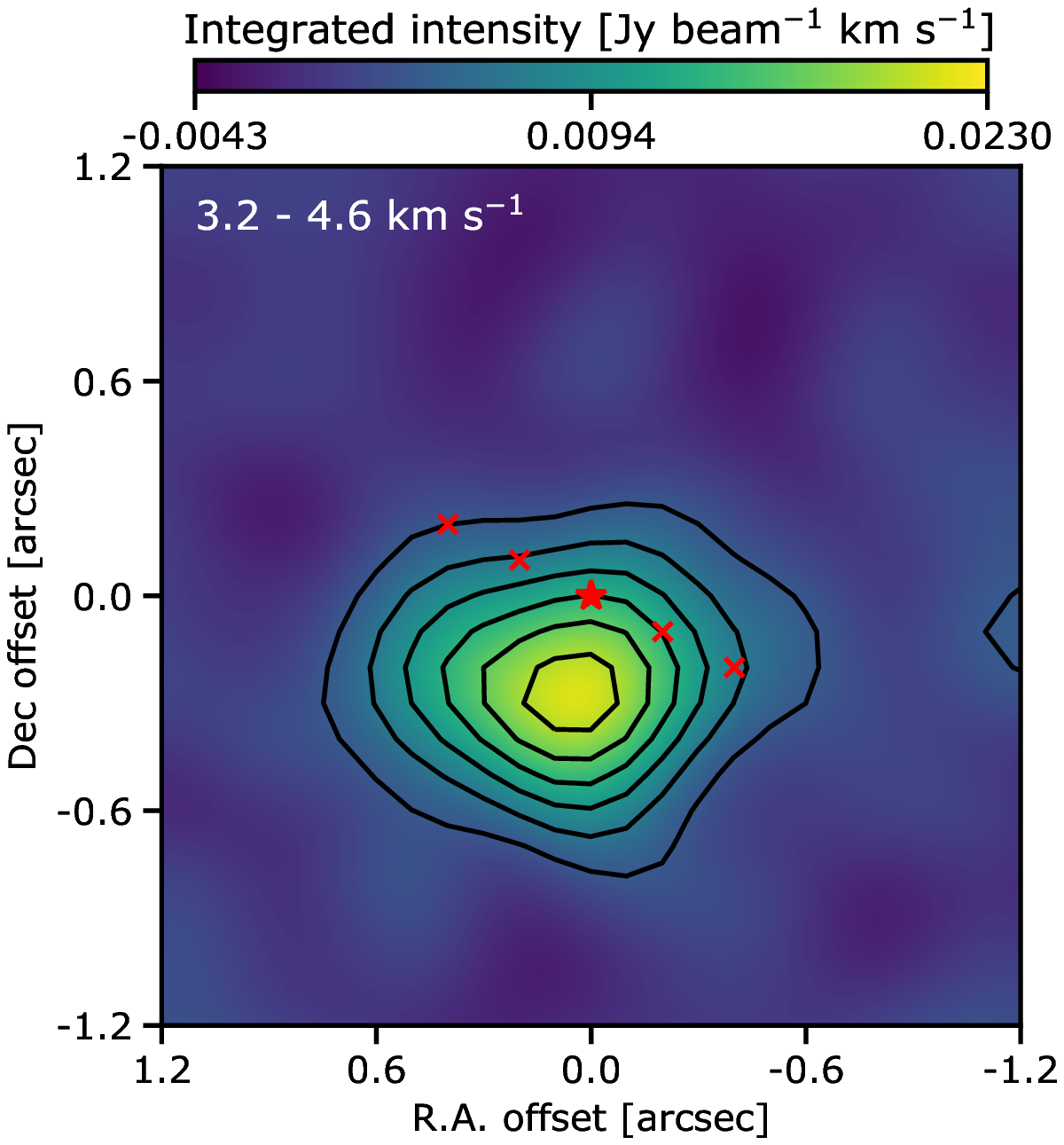} 
\includegraphics[width=0.6\columnwidth]{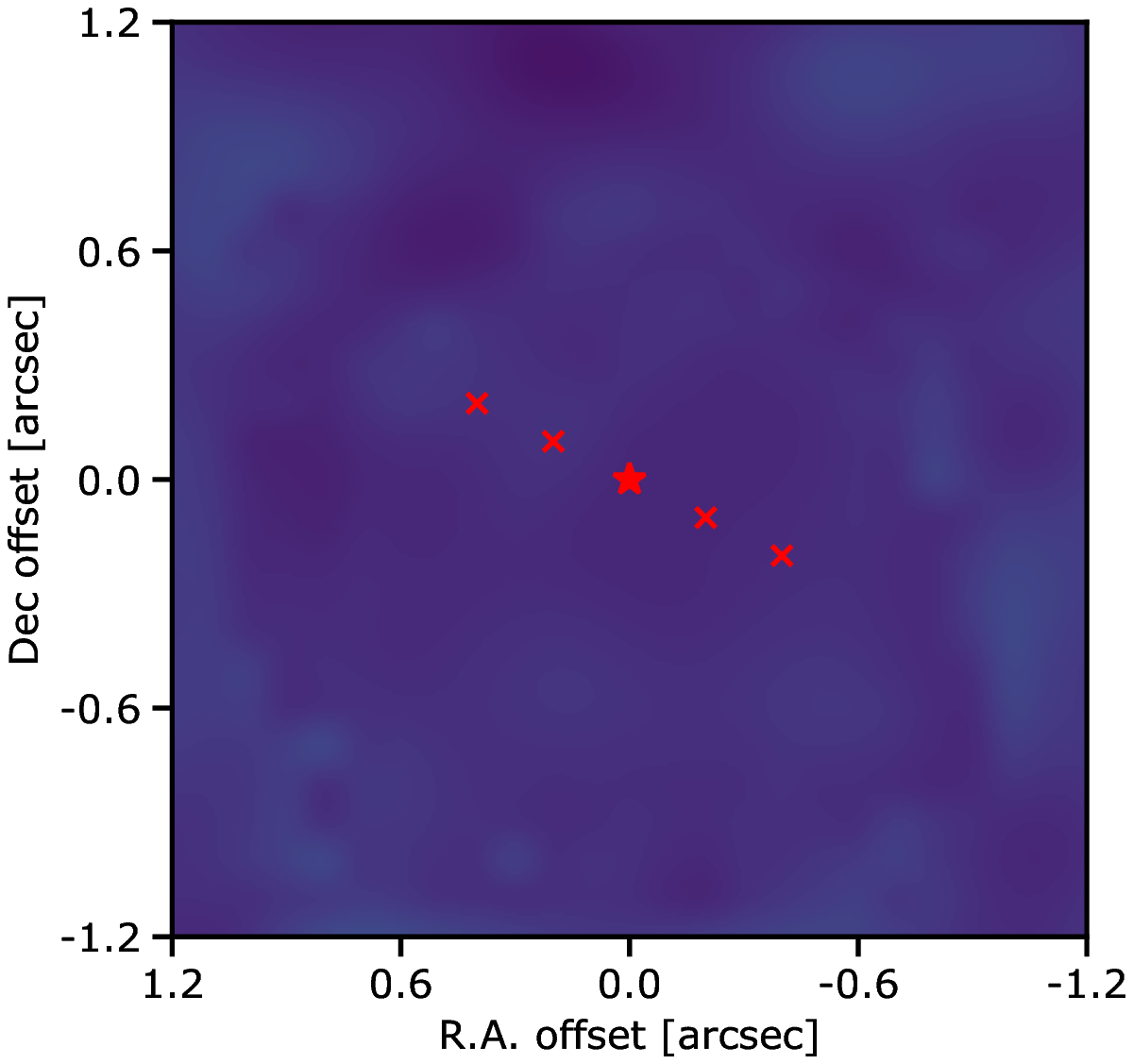}
\includegraphics[width=0.6\columnwidth]{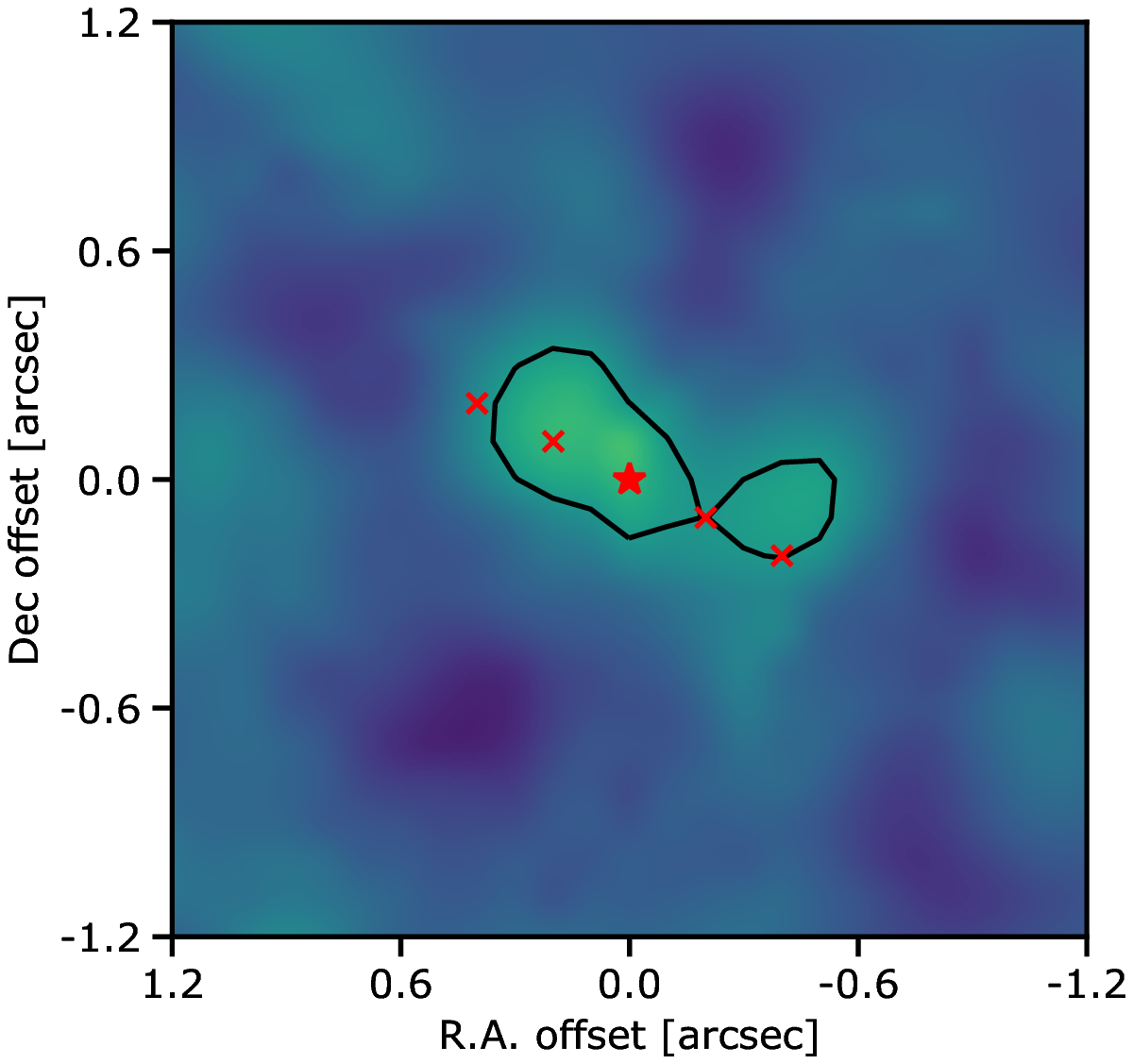}
\includegraphics[width=0.6\columnwidth]{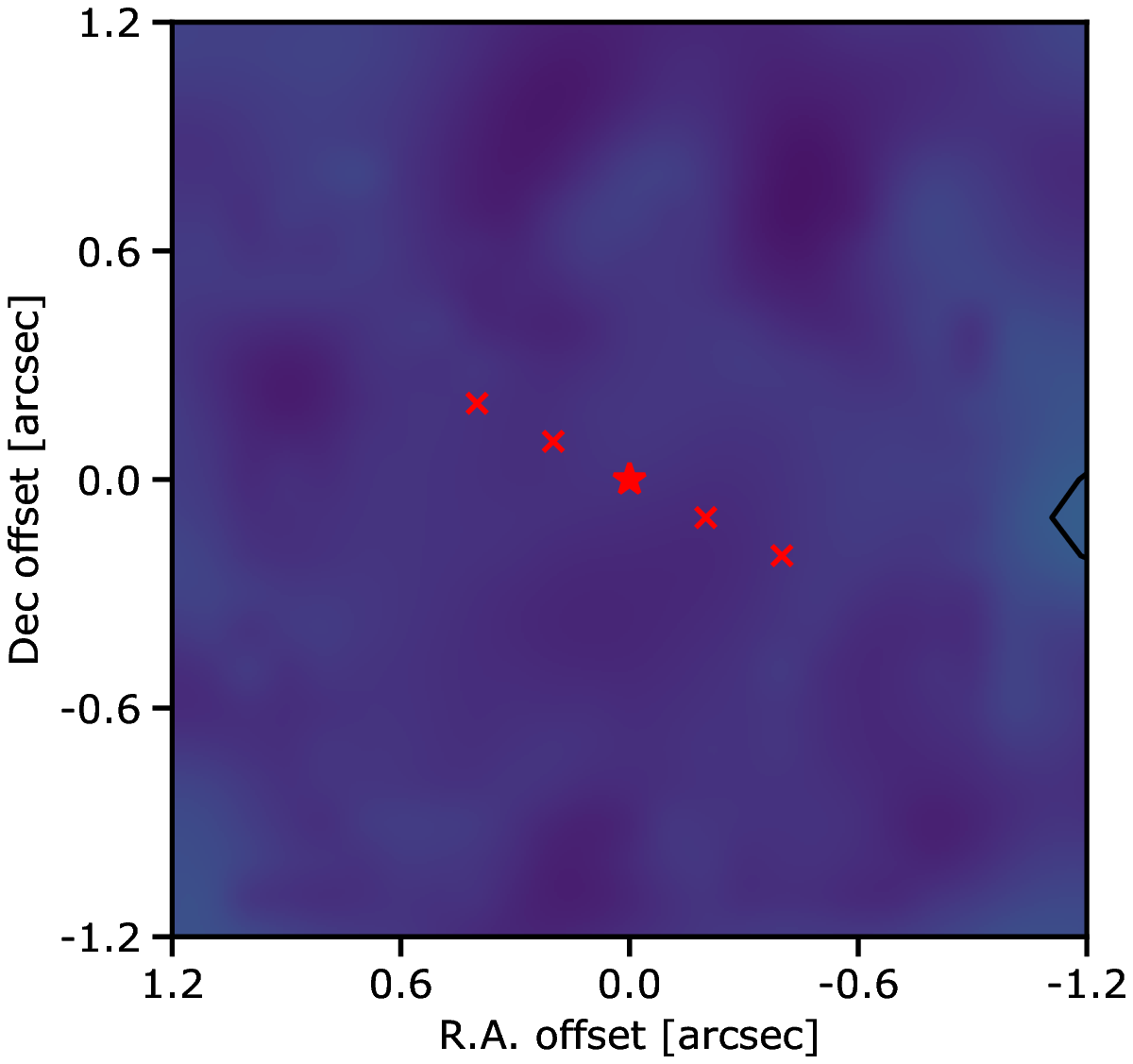}
\caption{Top: Integrated intensity maps around the $^{16}$O$^{18}$O transition at 233.946098 GHz for velocities of $1.3-2.8$ (left), $2.8-3.2$ (middle), and $3.2-4.6$ (right) km s$^{-1}$.
Bottom: Residual maps of the integrated intensity emission after subtraction of the fit performed with the line profile of the reference transition.
Contours are in steps of 3$\sigma$, with $\sigma$ of 1.07, 0.54, and 1.08 mJy beam$^{-1}$ km s$^{-1}$, respectively. The red star symbols depicts the position of the IRAS16293 B continuum peak while red crosses show the positions of the half-beam and full-beam offset positions mentioned in the text.
  }
\label{maps_o2}
\end{figure*}

The two surrounding transitions have first been fitted by Gaussian functions. Faint excess emission with respect to the Gaussian best-fits can be noticed between the two brighter surrounding transitions at $\sim$ 3.0 km s$^{-1}$, in particular for the continuum peak and the two western positions, as shown in the residual spectra in Fig. \ref{spect_o2}. 
However, line profiles around IRAS16293 B do not necessarly follow symmetric Gaussian profiles \citep[see][]{Zapata2013} and the weak intensity excess seen at $\sim 3.0$ km s$^{-1}$ could be due to the complex line profiles of the two surrounding transitions. 
We therefore used the profile of other nearby transitions as references to fit the profiles of the two contaminating transitions towards each pixel of the map around IRAS16293 B. We looked for nearby optically thin transitions with similar intensities and linewidths and free from contamination from other lines. The transition chosen as reference is the CH$_3$NCO transition at 234.08809 GHz but several other transitions with similar profiles could have been used. 
For each pixel of the datacube, we fitted the two transitions with the profile of the CH$_3$NCO transition by varying the intensity maximum and the intensity peak velocity when one of the transitions is detected above the 2$\sigma$ level. 
The spectra of the residual emission are shown in Fig. \ref{spect_o2} while the integrated emission maps before and after the subtraction of the best-fits are shown in Fig. \ref{maps_o2} for three velocity ranges: $1.3 - 2.8$, $2.8 - 3.2$, and $3.2 - 4.6$ km s$^{-1}$.
The residual spectra still show a weak intensity emission around $2.8 - 3.2$ km s$^{-1}$ towards the continuum peak and the two western positions both using the Gaussian and the "observed" line profiles. The intensity peaks are about $4-6$ mJy beam$^{-1}$, therefore just above the 3$\sigma$ limit with a rms noise of $\sim$ 1.3 mJy beam$^{-1}$ channel$^{-1}$. 
The weak emission can also be seen in the integrated emission maps in Fig. \ref{maps_o2}. Although no residual emission is detected at a 3$\sigma$ level for the surrounding transition velocity ranges, the residual map at $2.8-3.2$ km s$^{-1}$ shows some emission above the 3$\sigma$ level towards IRAS16293 B. 

Given the low signal-to-noise of the residual transition and its velocity shift with respect to the source velocity of 2.7 km s$^{-1}$, we derive the O$_2$ column density by considering a non detection and a tentative detection.
An upper limit to the O$_2$ column density is first obtained by deriving the 3$\sigma$ intensity upper limits $3 \sigma \sqrt{FWHM \delta v}$ of the transition at 233.946 GHz of the residual spectrum, where $\sigma$ is the rms noise of the spectrum, $FWHM$ is the expected FWHM linewidth of the transition, assumed to be 1.0 km s$^{-1}$, and $\delta v$ is the velocity resolution (= 0.156 km s$^{-1}$).
We assumed Local Thermodynamic Equilibrium (LTE) and we varied the excitation temperature $T_{\rm ex}$ between 125 K and 300 K, the excitation temperatures usually derived for other species near IRAS16293 B \citep[see][]{Lykke2017, Jorgensen2018}. We obtain an upper limit in $N$($^{16}$O$^{18}$O) of $(1.5 - 3.2) \times 10^{17}$ cm$^{-2}$, implying an upper limit in the O$_2$ column density of $(4.2 - 9.0) \times 10^{19}$ cm$^{-2}$, assuming a $^{16}$O$^{18}$O/O$_2$ abundance ratio of 280 taking into account that $^{18}$O can be in two positions in the molecule \citep{Wilson1994}. 


Assuming now that the residual transition is real and is due to the presence of $^{16}$O$^{18}$O, we derive its integrated intensity towards the western half-beam position peak through a Gaussian fit. We thus obtain a $^{16}$O$^{18}$O column density of $(3.5 - 7.5) \times 10^{17}$ cm$^{-2}$ for $T_{\rm ex} = 125 - 300$ K, giving $N$(O$_2$) $= (9.9 - 21) \times 10^{19}$ cm$^{-2}$. 

\begin{table*}[htp]
\centering
\caption{Column densities and abundances relative to CH$_3$OH of O$_2$, HO$_2$, and H$_2$O$_2$.} 
\begin{tabular}{l c c c}
\hline							
\hline							
Molecule    & $N$	  	      	      	          & \multicolumn{2}{c}{$N$/$N$(CH$_3$OH) (\%)}	\\
	        & (cm$^{-2}$) 	      	      	      & IRAS 16293	&	Comet 67P/C-G $^{\rm c}$	\\
\hline					
O$_2$	    &	$\leq (4.2 - 9.0) \times 10^{19}$ & $\leq 2.1 - 4.5$ $^{\rm a}$	        &	$5 - 15$     \\
	        &	$(9.9 - 21) \times 10^{19}$       & $5.0 - 10.5$ $^{\rm b}$	            &	$5 - 15$     \\
H$_2$O$_2$	&	$-$                               & $-$                                 &	$0.27 - 1.0$ \\
HO$_2$	    &	$\leq (1.1 - 2.8) \times 10^{14}$ & $\leq (5.5 - 14) \times 10^{-4}$	&	$0.85 - 3.2$ \\
\hline									
\end{tabular}
\label{transitions}
\tablebib{
 {  Column densities have been derived assuming excitation temperatures between 125 and 300 K (see text for more details). }
a: Abundance assuming that the $^{16}$O$^{18}$O transition is not detected.
b: Abundance assuming that the $^{16}$O$^{18}$O transition is detected.
c: Abundances derived following the abundances measured by \citet{Leroy2015} and \citet{Bieler2015}.
}
\end{table*}

\subsection{Analysis of the HO$_2$ and H$_2$O$_2$ transitions}

\begin{figure}[htp]
\centering 
\includegraphics[height=0.51\columnwidth]{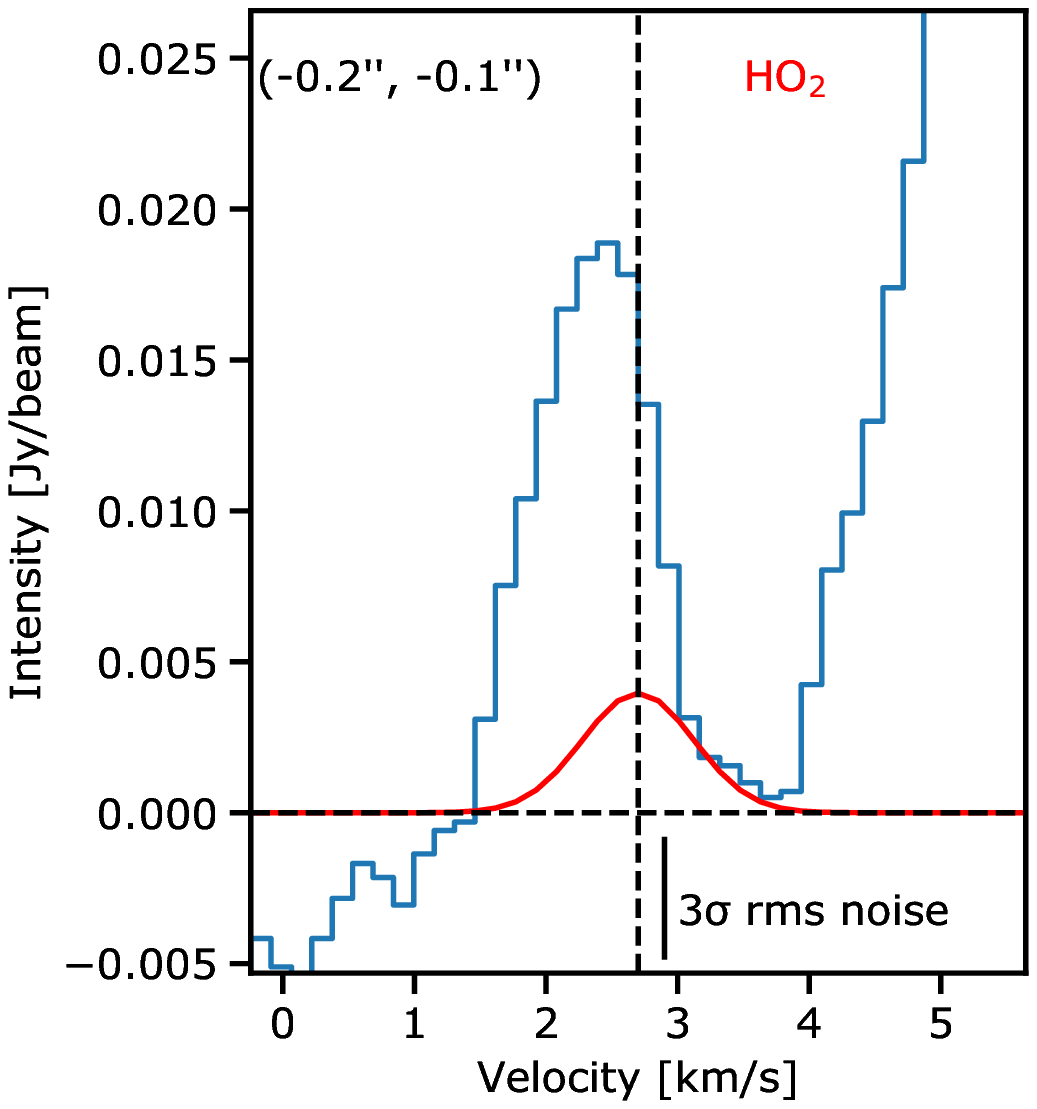} 
\includegraphics[height=0.51\columnwidth]{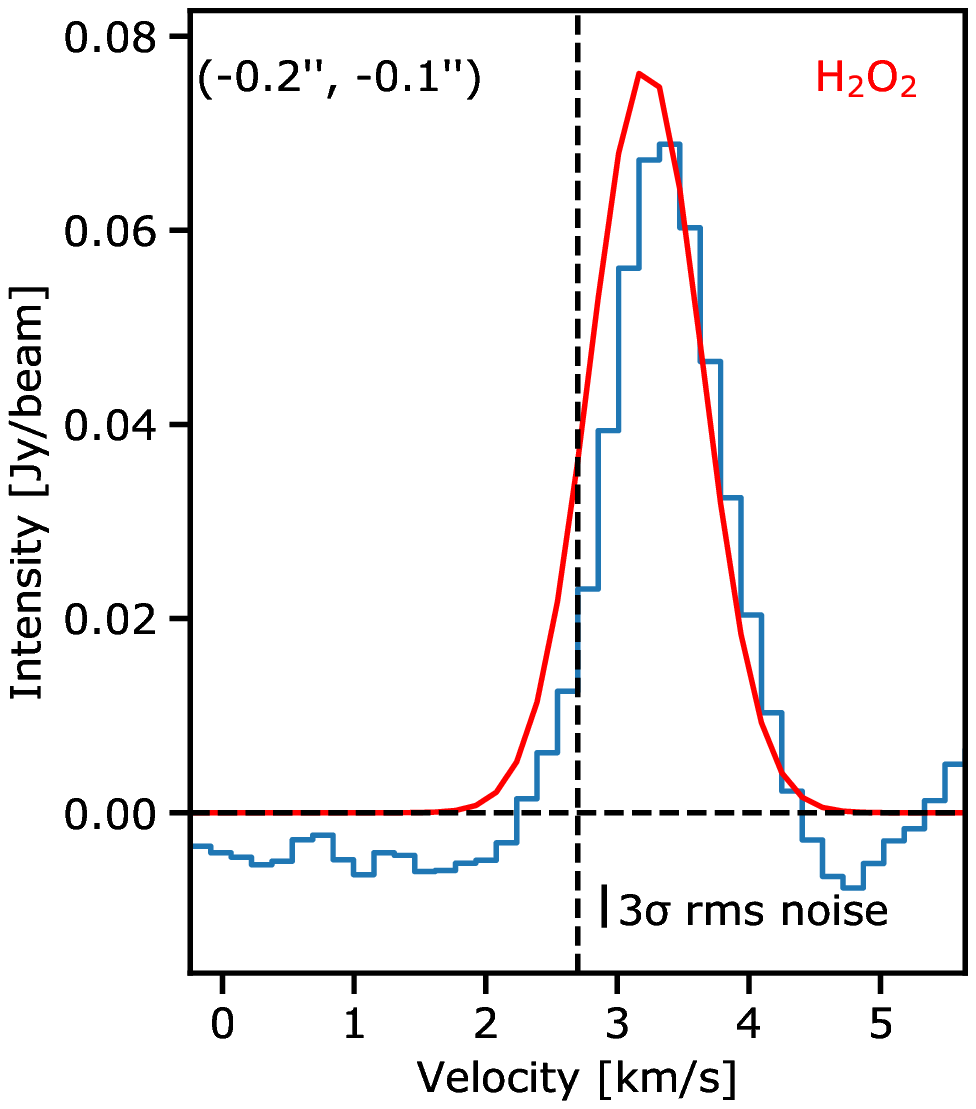} 
\caption{
Observed spectra (in blue) around the HO$_2$ transition at 236.280920 GHz (left panel) and around the H$_2$O$_2$ transition at 235.955943 GHz (right panel), respectively towards the western half beam position. 
The red line in the left and right panels shows the modelled spectrum at LTE obtained with a HO$_2$ column density of $1.1 \times 10^{14}$ cm$^{-2}$ and $T_{\rm ex} = 125$ K and a c-C$_2$H$_4$O column density of $6.1 \times 10^{15}$ cm$^{-2}$ and $T_{\rm ex}$ = 125 K respectively (see text). 
Positions in arcsec relative to the continuum peak position of IRAS16293 B are shown at the top left of each panel. 
The vertical and horizontal black dashed lines depict the source velocity at 2.7 km s$^{-1}$ and the baseline, respectively.
}
\label{spect_h2o2}
\end{figure}

Only one detectable H$_2$O$_2$ transition lies in our ALMA Band 6 dataset at 235.955 GHz. Unfortunately, the H$_2$O$_2$ transition is dominated by a transition from the ethylene oxide c-C$_2$H$_4$O species already detected in the Band 7 PILS data by \cite{Lykke2017}. 
Fig. \ref{spect_h2o2} compares the spectrum observed towards the western full-beam offset position around the H$_2$O$_2$ frequency with the synthetic spectrum assuming the column densities and excitation temperatures derived by \cite{Lykke2017}. 
The LTE model gives a reasonable fit to the observed transition suggesting that c-C$_2$H$_4$O is likely responsible for most, if not all, of the transition intensity. It is therefore impossible to conclude anything on the presence of H$_2$O$_2$ in IRAS16293 B because no detectable H$_2$O$_2$ transitions lie in the PILS Band 7 data.

The spectral windows have also been chosen to observe five bright transitions from HO$_2$ whose frequencies and properties are listed in Table \ref{transitions}. 
The transition at 235.170 GHz is contaminated by an ethyl glycol transition and the transition at 236.284 GHz is contaminated by a methyl formate transition. The LTE model presented in section \ref{overview_spectrum} gives a good fit to these transitions and do not allow us to use them to confirm the presence of HO$_2$.
None of the remaining transitions are detected. The transition at 236.280 GHz would have given the strongest constraint on the upper limit in HO$_2$ column density because of its high Einstein coefficient ($A_{\rm i,j} = 7.7 \times 10^{-5}$ s$^{-1}$). Figure \ref{spect_h2o2} shows the spectrum around the transition at 236.280 GHz towards the western full-beam offset position. 
It can be seen that a transition peaking at 2.4 km s$^{-1}$ is partially contaminating the targeted HO$_2$ transition. We have not been able to identify the species responsible for this transition. Ethyl formate, {\it trans}-C$_2$H$_5$OCHO was thought to be a plausible species since the frequency of two bright transitions match that of the observed line. However, this species has been ruled out because brighter transitions are not detected in our ALMA dataset.
In order to derive the upper limit in the HO$_2$ column density, we vary the column density that reproduces best the wing between 3.0 and 3.6 km s$^{-1}$ of the observed spectrum for excitation temperatures between 125 K and 300 K assuming LTE emission. We obtained upper limits of $N$(HO$_2$) $\leq 1.1 \times 10^{14}$ and $\leq 2.8 \times 10^{14}$ cm$^{-2}$ for $T_{\rm ex} = 125$ and 300 K, respectively. We confirmed {\it a posteriori} that the other HO$_2$ transitions are not detected at the 3$\sigma$ limit with the derived column densities.  

\section{Discussion and conclusions}

The $^{16}$O$^{18}$O transition at 233.946 GHz is contaminated by two brighter transitions at $\pm 1$ km s$^{-1}$ relative to the expected targeted transition frequency. After subtraction of these two transitions, residual emission remains at a 3$\sigma$ level but with a velocity offset of $0.3 - 0.5$ km s$^{-1}$ with respect to the source velocity. We therefore assume two cases, a tentative detection of the $^{16}$O$^{18}$O transition and a more realistic non-detection. In the following, we consider the non-detection case to compare the abundance of O$_2$ with water and methanol.
The H$_2$O abundance towards IRAS16293 B is still unknown because only one transition of H$_2^{18}$O has been detected in absorption by \citet{Persson2013} using ALMA. We therefore decide to use first methanol as a reference species. The CH$_3$OH column density $N$(CH$_3$OH) has been accurately derived using a large number of transitions from its optically thin CH$_3^{18}$OH isotopologue by \citet{Jorgensen2016, Jorgensen2018}, giving $N$(CH$_3$OH) $= 2 \times 10^{19}$ and $1 \times 10^{19}$ cm$^{-2}$ towards the western half-beam and full-beam offset positions of IRAS16293 B, respectively.  
For the case that the O$_2$ transition is a non-detection, we therefore derive [O$_2$]/[CH$_3$OH] $\leq 2.1 - 4.5$. 
%
H$_2$O ice has a typical abundance of $1 \times 10^{-4}$ relative to H$_2$ in molecular clouds \citep{Tielens1991, Pontoppidan2004, Boogert2015} and is expected to fully sublimate in the warm gas around protostars once the temperature exceeds the water sublimation temperature of $\sim 100$ K. 
\citet{Jorgensen2016} derived a lower limit for the H$_2$ column density $N$(H$_2$) $> 1.2 \times 10^{25}$ cm$^{-2}$, resulting in a lower limit in the H$_2$O column density $N$(H$_2$O) $> 1.2 \times 10^{21}$ cm$^{-2}$. This results in estimates of [CH$_3$OH]/[H$_2$O] $< 1.6$\% and [O$_2$]/[H$_2$O] $< 3.5$\%. 

Gaseous O$_2$ detected in the hot core of IRAS16293 B is expected to result mostly from the sublimation of solid O$_2$ locked into ices \citep[see][for instance]{Yildiz2013}.
Gas-phase "hot-core" chemistry that could destroy O$_2$, through UV photo-dissociation or neutral-neutral reactions, after its evaporation from the interstellar ices is likely inefficient. 
Photo-dissociation is probably not at work here due to the optically thick protostellar envelope present around the young Class 0 protostar that shields any strong UV radiation field. 
In addition, the gas phase temperature around IRAS16293 B of $100 - 300$ K is likely too low to trigger the reactivity of the reaction between O$_2$ and H because of the large activation barrier of 8380 K. 
The O$_2$ abundance inferred from our ALMA observations should therefore reflect the abundance of icy O$_2$ in the prestellar core at the origin of the IRAS16293 protostellar system. 

The O$_2$ abundance lower than $3.5$\% relative to water is consistent with the upper limits in the solid O$_2$ abundance of $\leq 15$ and $\leq 39$\% relative to H$_2$O found towards the low-mass protostar R CrA IRS2 and the massive protostar NGC 7538 IRS9, respectively \citep{Vandenbussche1999, Boogert2015}. 
The upper limits are also consistent with the predictions by \citet{Taquet2016} who modelled the formation and survival of solid O$_2$ for a large range of dark cloud conditions and found values lower than a few percents for a large range of model parameters. The extended core of IRAS16293 has a dust temperature of 16 K, based on observations of the Ophiuchus cloud with the SPIRE and PACS instruments of the {\it Herschel Space Observatory} as part of the Gould Belt Survey key program \citep[][B. Ladjelate, private communication]{Andre2010}. 
According to the \citet{Taquet2016} model predictions, assuming a temperature of 16 K, the prestellar core that formed IRAS16293 could have spent most of its lifetime at a density lower than $10^5$ cm$^{-3}$ and/or a cosmic ray ionisation rate higher than $10^{-17}$ s$^{-1}$, allowing ice formation with an efficient hydrogenation process that favours the destruction of O$_2$ into H$_2$O$_2$ and H$_2$O. 


As the O$_2$ abundance derived around IRAS16293 B likely reflects the O$_2$ abundance in interstellar ices before their evaporation, it can be compared with the abundance measured in comets to follow the formation and survival of O$_2$ from dark clouds to planetary systems.
CH$_3$OH has also been detected in comet 67P/C-G at the same mass 32 as O$_2$ by the ROSINA mass spectrometer onboard {\it Rosetta} with an abundance of $0.31-0.55$ \% relative to H$_2$O \citep{Leroy2015}, implying a [O$_2$]/[CH$_3$OH] abundance ratio of $5.3 - 15$.
Under the safer assumption that O$_2$ is not detected towards IRAS1693 B, the derived upper limit [O$_2$]/[CH$_3$OH] $\leq 2.1 - 4.5$ measured in IRAS16293 B is slightly lower than the abundance measured in 67P/C-G. 
However, the CH$_3$OH abundance is about 10 times lower in 67P/C-G than the median abundance found in interstellar ices towards a sample of low-mass protostars \citep[CH$_3$OH/H$_2$O $\sim 7$\%,][]{Bottinelli2010, Oberg2011}. The low CH$_3$OH abundance measured in comet 67P/C-G could explain the differences in [O$_2$]/[CH$_3$OH] between IRAS16293 and 67P. 
Using water as a reference species, the [O$_2$]/[H$_2$O] $\leq 3.5$\% in IRAS16293 B falls within the abundance range of 2.95 - 4.65 \% observed in comet 67P/C-G by ROSINA. 
With a temperature of 16 K, the precursor dark cloud of IRAS16293 is slightly colder than the temperature of $20-25$ K required to enhance the O$_2$ formation in interstellar ices within dark clouds \citep{Taquet2016}. 
Further interferometric observations of $^{16}$O$^{18}$O towards other bright nearby low-mass protostars located in warmer environments than the cloud surrounding IRAS16293 could result in an unambiguous detection of molecular oxygen O$_2$ around young protostars.
Such a study would confirm the primordial origin of cometary O$_2$ in our Solar System.


\begin{acknowledgements}
This paper makes use of the following ALMA data: ADS/JAO.ALMA\#2016.1.01150.S. 
ALMA is a partnership of ESO (representing its member states), NSF (USA) and NINS (Japan), together with NRC (Canada) and NSC and ASIAA (Taiwan), in cooperation with the Republic of Chile. The Joint ALMA Observatory is operated by ESO, AUI/NRAO and NAOJ. 
V.T. acknowledges the financial support from the European Union's Horizon 2020 research and innovation programme under the Marie Sklodowska-Curie grant agreement n. 664931.
Astrochemistry in Leiden is supported by the European Union A-ERC grant 291141 CHEMPLAN, by the Netherlands Research School for Astronomy (NOVA) and by a Royal Netherlands Academy of Arts and Sciences (KNAW) professor prize.
J.K.J. acknowledges support from the European Research Council (ERC) under the European Union's Horizon 2020 research and innovation programme (grant agreement No~646908) through ERC Consolidator Grant ``S4F''
A.C. postdoctoral grant is funded by the ERC Starting Grant 3DICE (grant agreement 336474).
C.W. acknowledges financial support from the University of Leeds.

\end{acknowledgements}

\appendix

\section{Spectra of the full spectral windows}

\begin{figure*}[htp]
\centering 
\includegraphics[width=\textwidth]{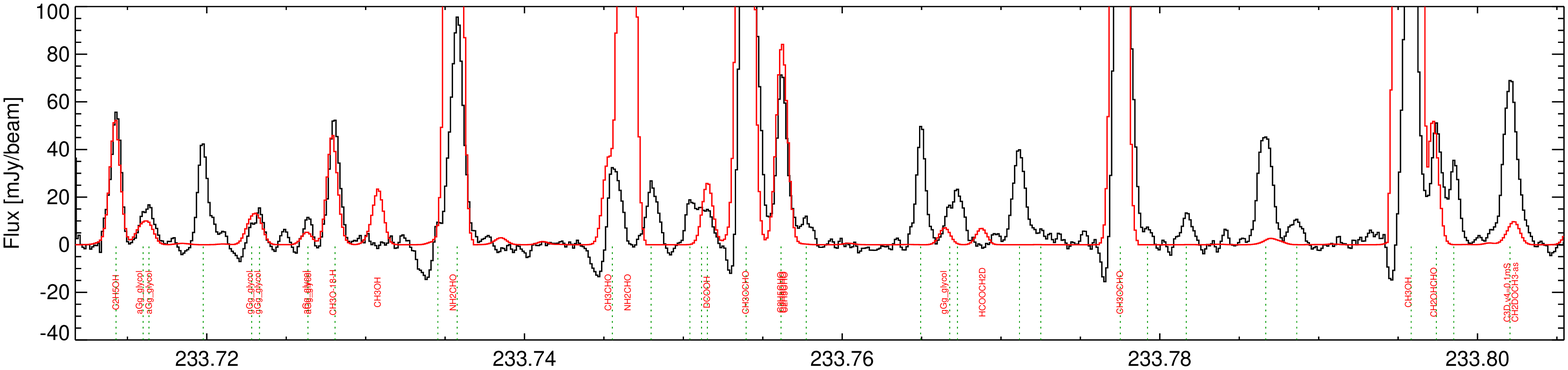}
\includegraphics[width=\textwidth]{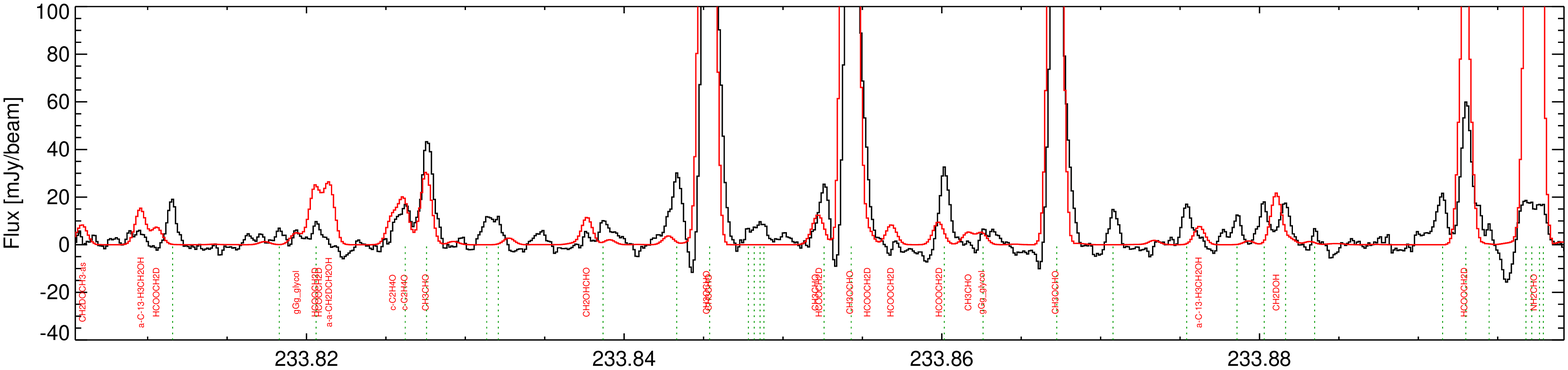}
\includegraphics[width=\textwidth]{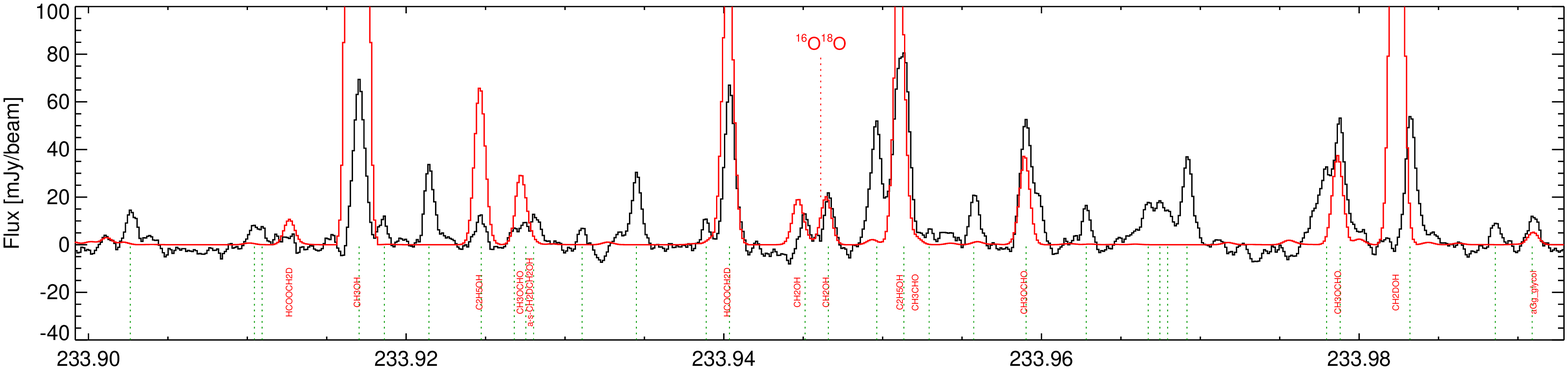}
\includegraphics[width=\textwidth]{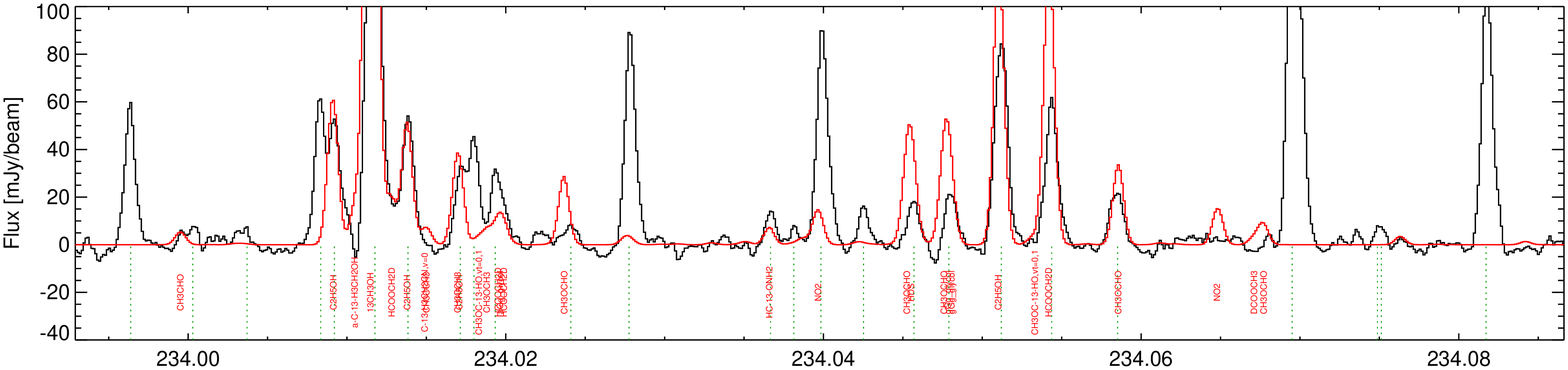}
\includegraphics[width=\textwidth]{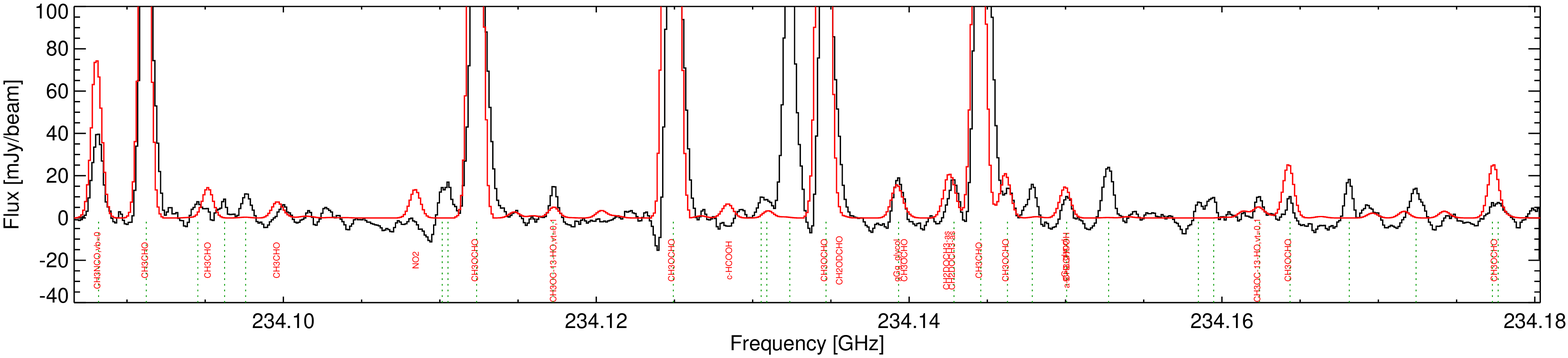}
\caption{
Spectrum (black) of the first spectral window between 233.712 and 234.180 GHz obtained towards the full-beam offset position located 0.5$\arcsec$ away from the continuum peak of IRAS16293 B in the south-west direction.
Synthetic spectrum of the LTE model is shown in red (see text for more details). {  Green dotted lines refer to the position of transitions of unidentified species detected above 5$\sigma$. }
  }
\label{spect_spw1}
\end{figure*}

\begin{figure*}[htp]
\centering 
\includegraphics[width=\textwidth]{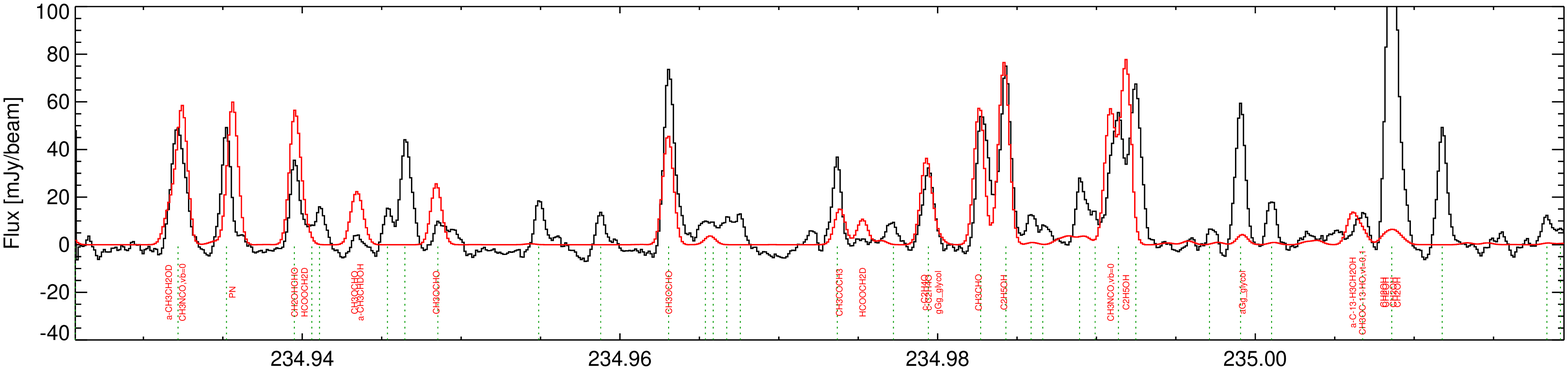}
\includegraphics[width=\textwidth]{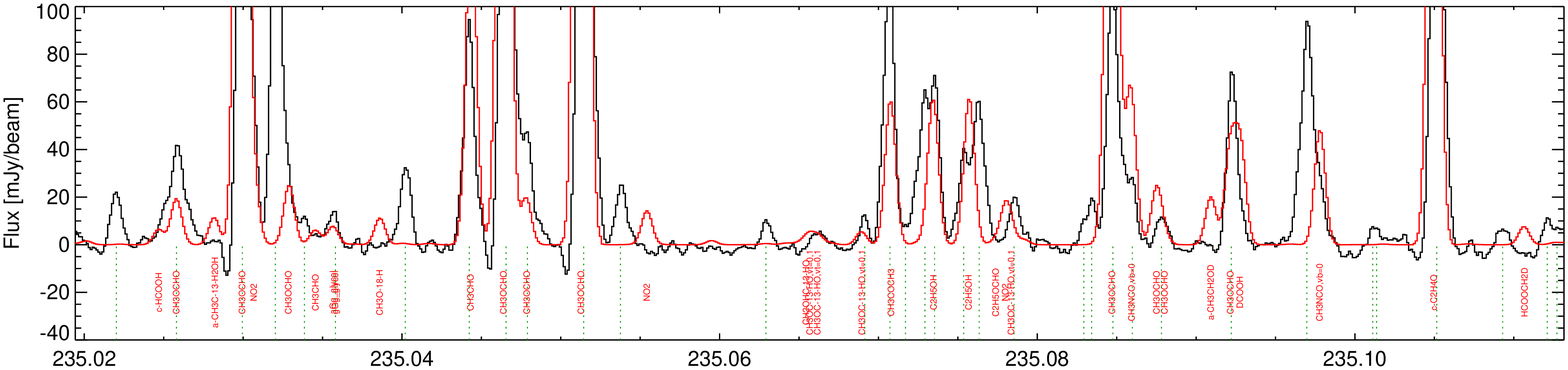}
\includegraphics[width=\textwidth]{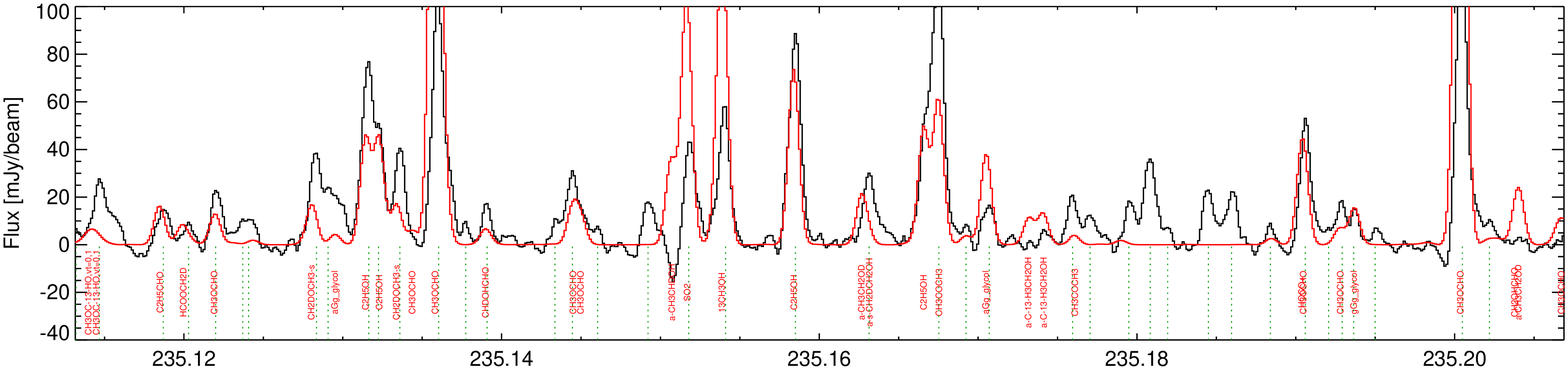}
\includegraphics[width=\textwidth]{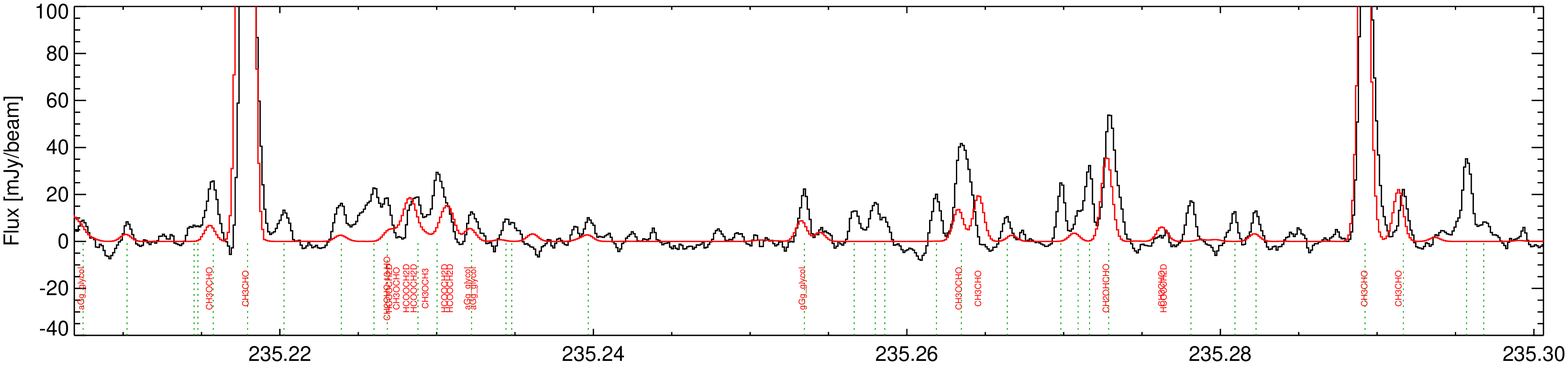}
\includegraphics[width=\textwidth]{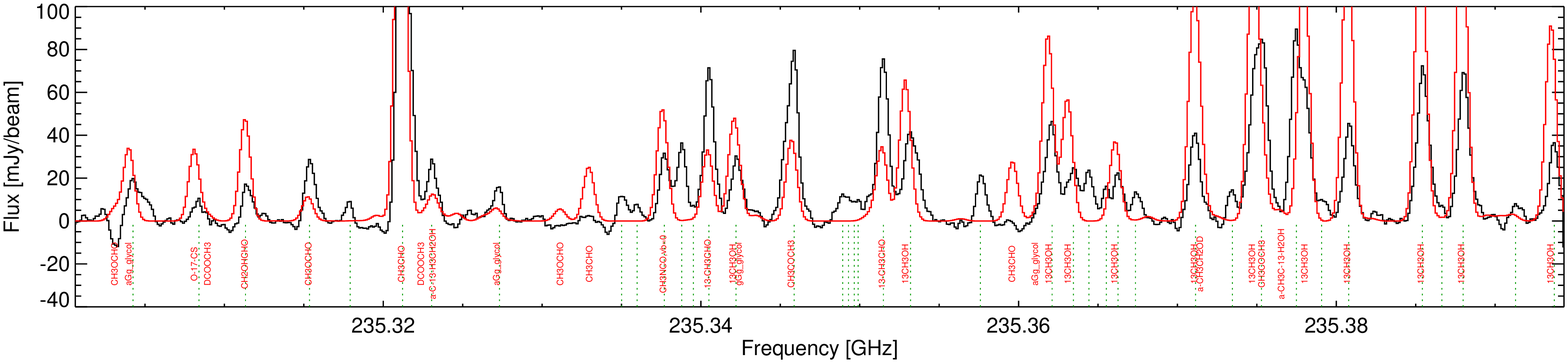}
\caption{
Same as Fig. \ref{spect_spw1} but for the spectral window between 234.918 and 235.385 GHz.
  }
\label{spect_spw2}
\end{figure*}

\begin{figure*}[htp]
\centering 
\includegraphics[width=\textwidth]{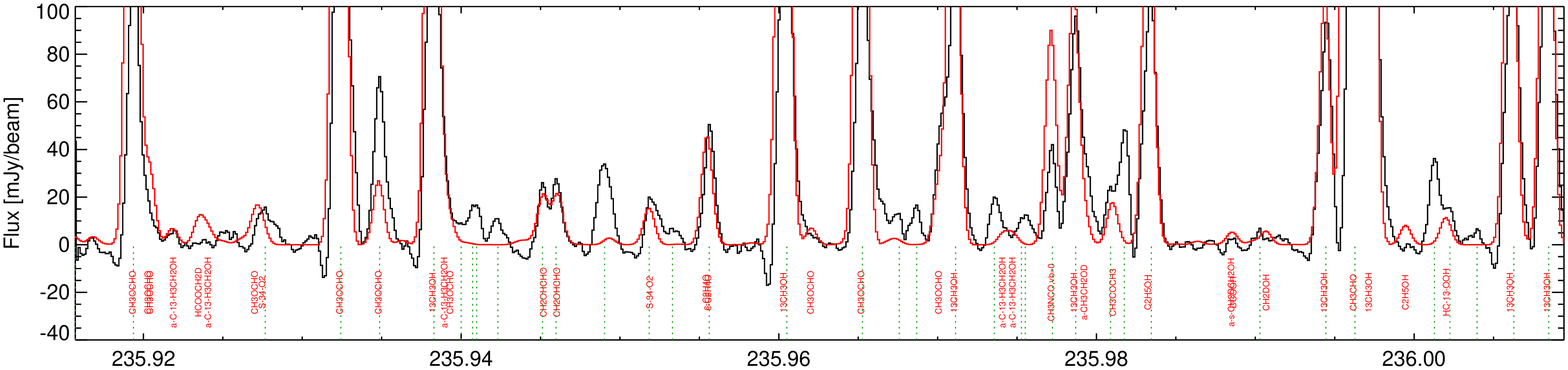}
\includegraphics[width=\textwidth]{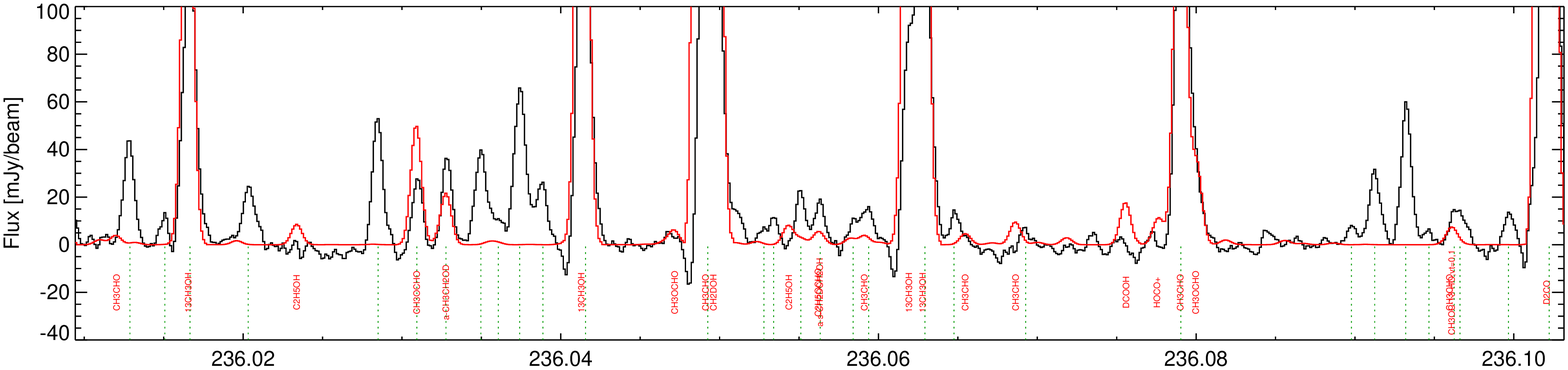}
\includegraphics[width=\textwidth]{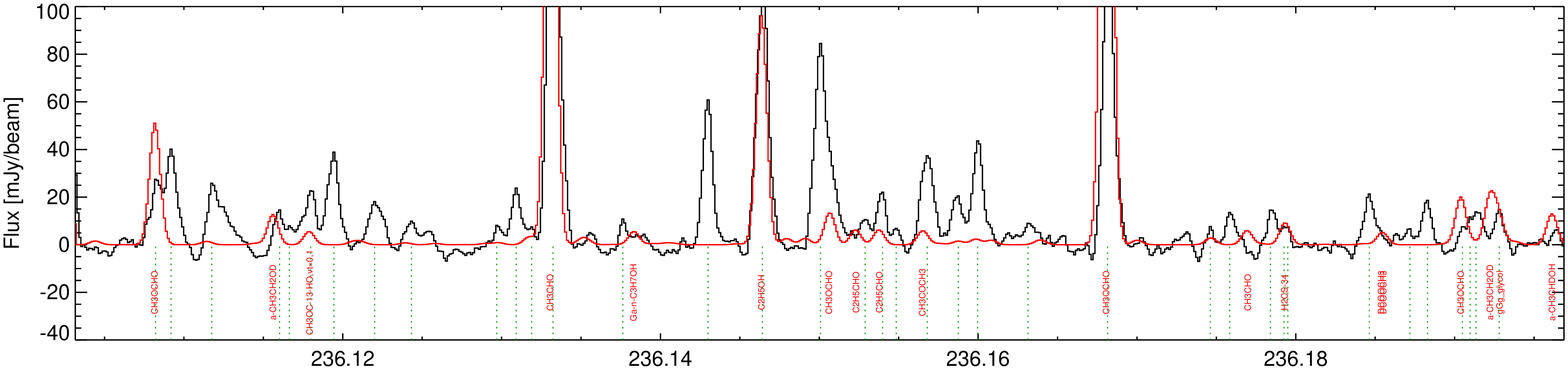}
\includegraphics[width=\textwidth]{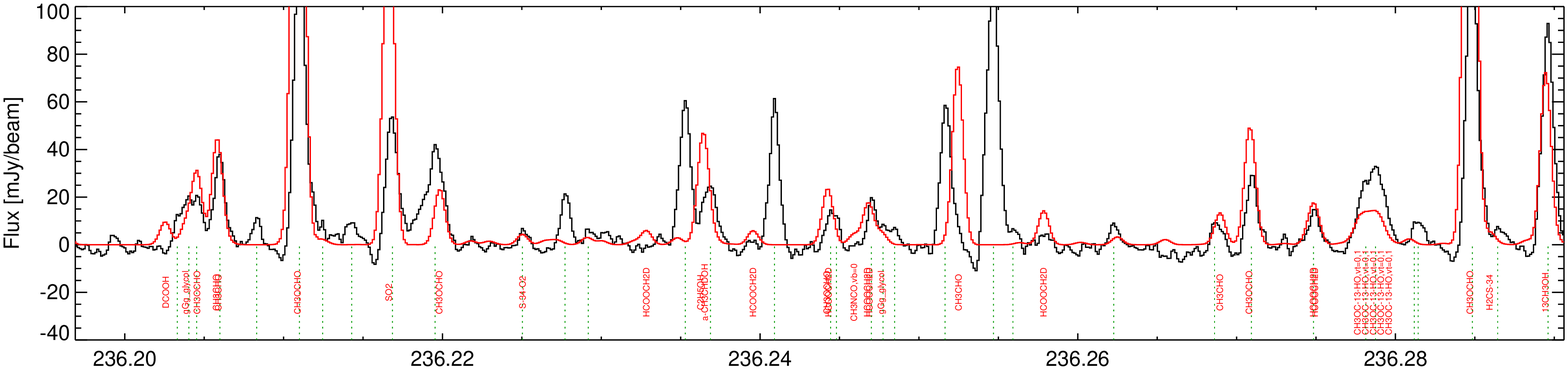}
\includegraphics[width=\textwidth]{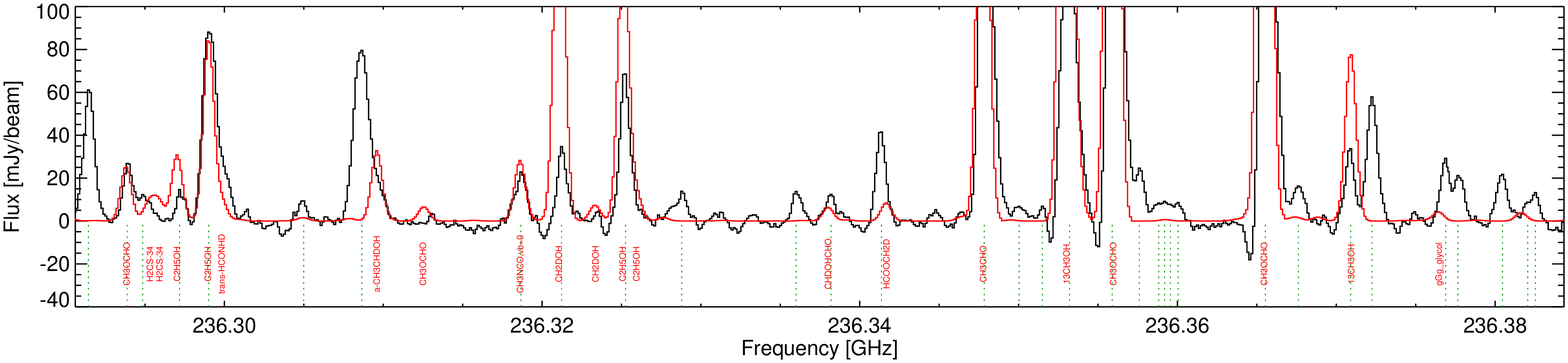}
\caption{
Same as Fig. \ref{spect_spw1} but for the spectral window between 235.908 and 236.379 GHz.
  }
\label{spect_spw3}
\end{figure*}

\begin{figure*}[htp]
\centering 
\includegraphics[width=\textwidth]{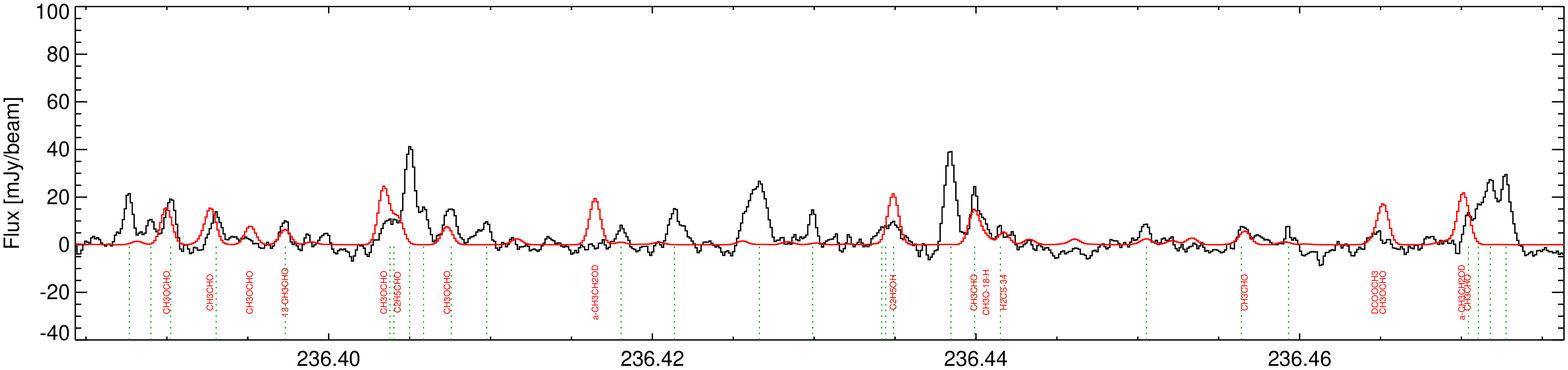}
\includegraphics[width=\textwidth]{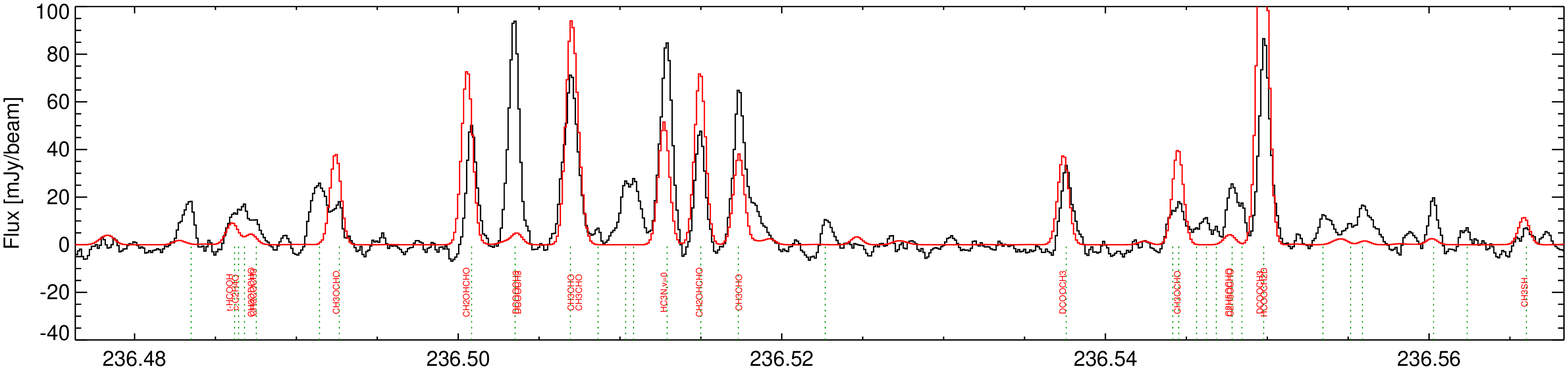}
\includegraphics[width=\textwidth]{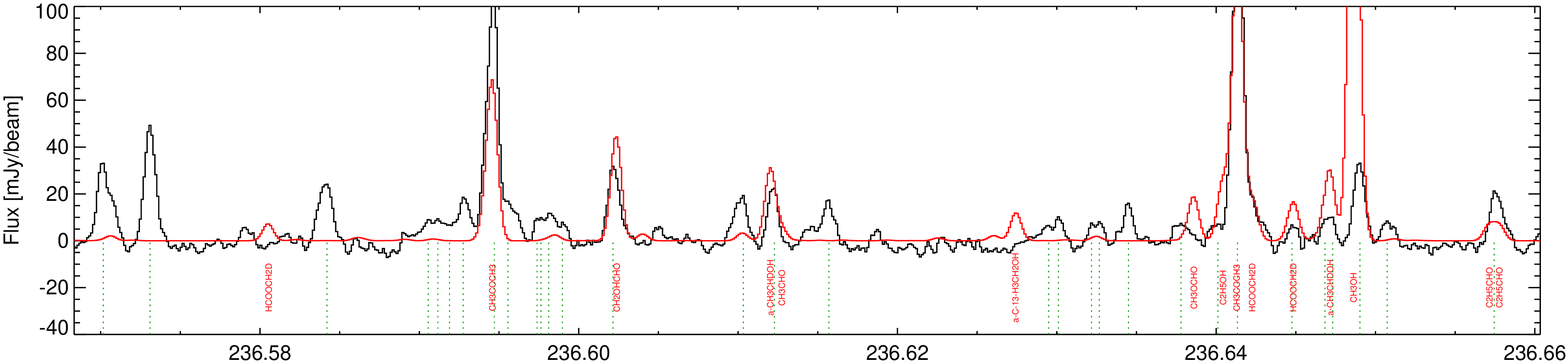}
\includegraphics[width=\textwidth]{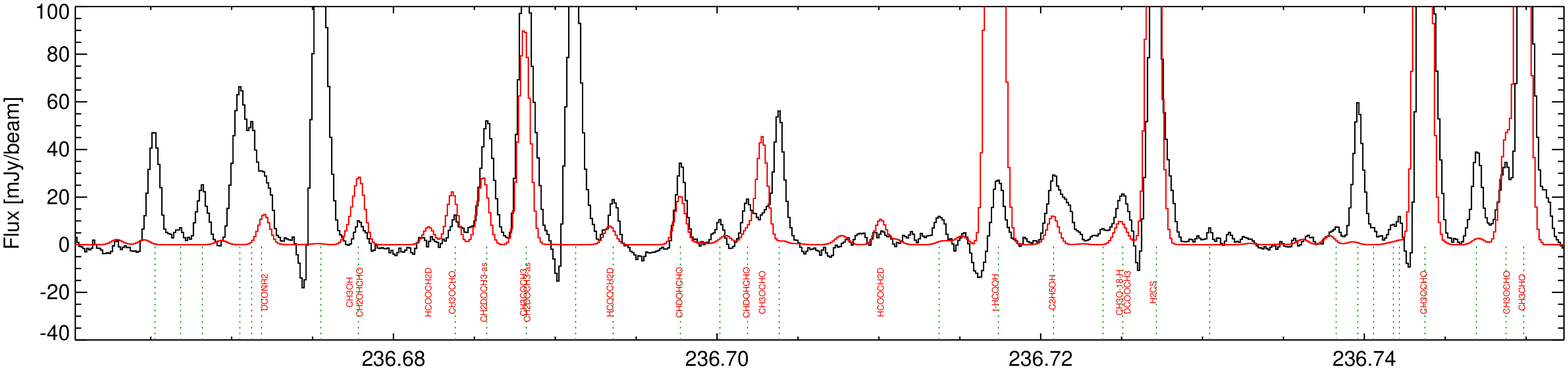}
\includegraphics[width=\textwidth]{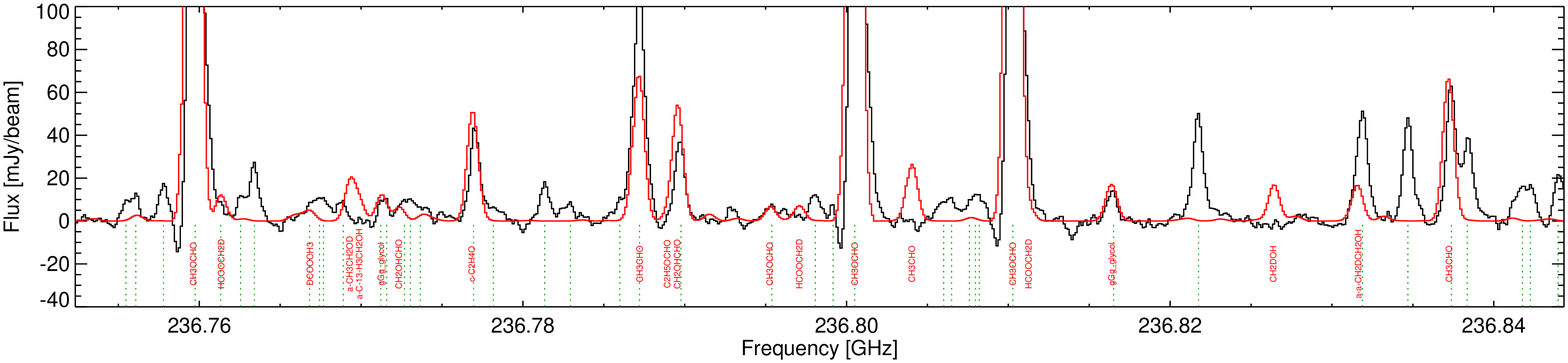}
\caption{
Same as Fig. \ref{spect_spw1} but for the spectral window between 236.379 and 236.841 GHz.
  }
\label{spect_spw4}
\end{figure*}

\end{document}